\documentclass[journal=nalefd,manuscript=letter]{achemso}

\usepackage{chemformula} 
\usepackage[T1]{fontenc} 
\usepackage{times}
\usepackage{upgreek}
\usepackage{float}
\usepackage{amsmath}
\usepackage{graphicx}
\usepackage{hyperref}
\usepackage{bibunits}
\usepackage{titlesec}
\SectionNumbersOn
\defaultbibliography{Bibliography.bib}

\author{Francesco Borsoi}
\affiliation[Delft University of Technology]
{QuTech and Kavli Institute of Nanoscience, Delft University of Technology, 2600 GA Delft, The Netherlands}
\email{f.borsoi@tudelft.nl}

\author{Grzegorz P. Mazur}
\affiliation[Delft University of Technology]
{QuTech and Kavli Institute of Nanoscience, Delft University of Technology, 2600 GA Delft, The Netherlands}

\author{Nick van Loo}
\affiliation[Delft University of Technology]
{QuTech and Kavli Institute of Nanoscience, Delft University of Technology, 2600 GA Delft, The Netherlands}

\author{Micha\l{} P. Nowak}
\affiliation[AGH University of Science and Technology]
{Academic Centre for Materials, AGH University of Science and Technology, al.\ A.\ Mickiewicza 30, 30-059 Krak\'ow, Poland}

\author{L\'eo Bourdet}
\affiliation[Delft University of Technology]
{QuTech and Kavli Institute of Nanoscience, Delft University of Technology, 2600 GA Delft, The Netherlands}

\author{Kongyi Li}
\affiliation[Delft University of Technology]
{QuTech and Kavli Institute of Nanoscience, Delft University of Technology, 2600 GA Delft, The Netherlands}

\author{Svetlana Korneychuk}
\affiliation[Delft University of Technology]
{QuTech and Kavli Institute of Nanoscience, Delft University of Technology, 2600 GA Delft, The Netherlands}

\author{Alexandra Fursina}
\affiliation[Microsoft Quantum Lab Delft]
{Microsoft Quantum Lab Delft, 2600 GA Delft, The Netherlands}

\author{Elvedin Memisevic}
\affiliation[Delft University of Technology]
{QuTech and Kavli Institute of Nanoscience, Delft University of Technology, 2600 GA Delft, The Netherlands}

\author{Ghada Badawy}
\affiliation[Eindhoven University of Technology]
{Department of Applied Physics, Eindhoven University of Technology, 5600 MB Eindhoven, The Netherlands}

\author{Sasa Gazibegovic}
\affiliation[Eindhoven University of Technology]
{Department of Applied Physics, Eindhoven University of Technology, 5600 MB Eindhoven, The Netherlands}

\author{Kevin van Hoogdalem}
\affiliation[Microsoft Quantum Lab Delft]
{Microsoft Quantum Lab Delft, 2600 GA Delft, The Netherlands}

\author{Erik P. A. M.~Bakkers}
\affiliation[Eindhoven University of Technology]
{Department of Applied Physics, Eindhoven University of Technology, 5600 MB Eindhoven, The Netherlands}

\author{Leo P. Kouwenhoven}
\affiliation[Delft University of Technology]
{QuTech and Kavli Institute of Nanoscience, Delft University of Technology, 2600 GA Delft, The Netherlands}
\alsoaffiliation[Microsoft Quantum Lab Delft]
{Microsoft Quantum Lab Delft, 2600 GA Delft, The Netherlands}

\author{Sebastian Heedt}
\affiliation[Microsoft Quantum Lab Delft]
{Microsoft Quantum Lab Delft, 2600 GA Delft, The Netherlands}

\author{Marina Quintero-P\'erez}
\affiliation[Microsoft Quantum Lab Delft]
{Microsoft Quantum Lab Delft, 2600 GA Delft, The Netherlands}
\email{marina.quintero@microsoft.com}

\title{Single-shot fabrication of semiconducting-superconducting nanowire devices}


\begin{document}


\newpage
\begin{abstract}
\noindent Semiconducting-superconducting nanowires attract widespread interest owing to the possible presence of non-abelian Majorana zero modes, which hold promise for topological quantum computation. However, the search for Majorana signatures is challenging because reproducible hybrid devices with desired nanowire lengths and material parameters need to be reliably fabricated to perform systematic explorations in gate voltages and magnetic fields.
Here, we exploit a fabrication platform based on shadow walls that enables the in-situ, selective and consecutive depositions of superconductors and normal metals to form normal-superconducting junctions. 
Crucially, this method allows to realize devices in a single shot, eliminating fabrication steps after the synthesis of the fragile semiconductor/superconductor interface. 
At the atomic level, all investigated devices reveal a sharp and defect-free semiconducting-superconducting interface and, correspondingly, we measure electrically a hard induced superconducting gap. While our advancement is of crucial importance for enhancing the yield of complex hybrid devices, it also offers a straightforward route to explore new material combinations for hybrid devices.
\end{abstract}

\begin{bibunit}
\pagebreak
Semiconductor-superconductor nanowires are prime candidates towards topological quantum computation based on the manipulation of Majorana zero modes~\cite{Oreg2010,Lutchyn2010,Alicea2012}. 
However, to serve as basic units of complex architectures~\cite{Hyart2013,Plugge2017,Karzig2017,Vijay2016}, hybrid nanowires require a homogeneous and pristine interface between the semi- and the superconductor.
In state-of-the-art methods, a superconducting film is deposited in-situ after the growth of the semiconductor nanowires~\cite{Krogstrup2015,Bjergfeld2019}, which are then transferred onto insulating substrates for further fabrication. The superconductor is chemically etched away from certain sections of the wires to realize gate-tunable regions.
This approach has a major drawback: the selectivity of the metal etching is often uncontrollable and results in damage to the semiconductor crystal, as well as in chemical contaminations~\cite{deMoor2019,Khan2020}.
An alternative method to obtain gate-tunable regions is the shadow evaporation of the superconductor. This can be obtained by engineering complex nanowire growth chips with trenches or horizontal bridges~\cite{Gazibegovic2017,Carrad2019,Khan2020}.
In this case, the nanowire growth needs to be accurately optimized, the variety of possible devices is minimal, and hybrid nanowires still need to be transferred onto a substrate and subsequently processed. 
Crucially, the semiconductor-superconductor interface is unstable and prone to degradation with time and temperature, a problem that is particularly severe for the case of InSb/Al where the degradation takes place even at room temperature~\cite{Pei1988,Boscherini1987,Sporken1988,Thomas2019,Gazibegovic2019}. The interface instability poses a limit to the development and systematic exploration of topological circuits.\\
\indent Here, we establish a fabrication method based on our previously introduced shadow-wall lithography technique~\cite{Heedt2020} that overcomes these problems and enables the synthesis of high-quality hybrid devices in a single shot. The complete elimination of processing after the formation of the delicate semiconductor-superconductor interface is the critical aspect of our approach.
To achieve this, we employ chips with pre-patterned bonding pads, bottom gates and shadow walls, next to which we transfer the semiconducting nanowires. The shadow walls and their particular design facilitate the selective deposition of the superconductor as well as contact leads without breaking the vacuum, eliminating the need for extra lithography steps. 
We demonstrate the versatility of our approach by creating hybrid junctions, which are the primary devices utilized to verify the emergence of Majorana excitations. 
The high quality of the devices is probed by transmission electron microscopy and by quantum transport measurements. 
All the investigated devices reveal a sharp and defect-free semiconductor-superconductor interface and, consequently, a hard induced superconducting gap.
Remarkably, our technique is generic. It can have vast applications, from sparking rapid exploration of different combinations of semiconductors and superconductors to paving the route to more complex devices.\\
\begin{figure}[hbt!]
	\centering
	\includegraphics{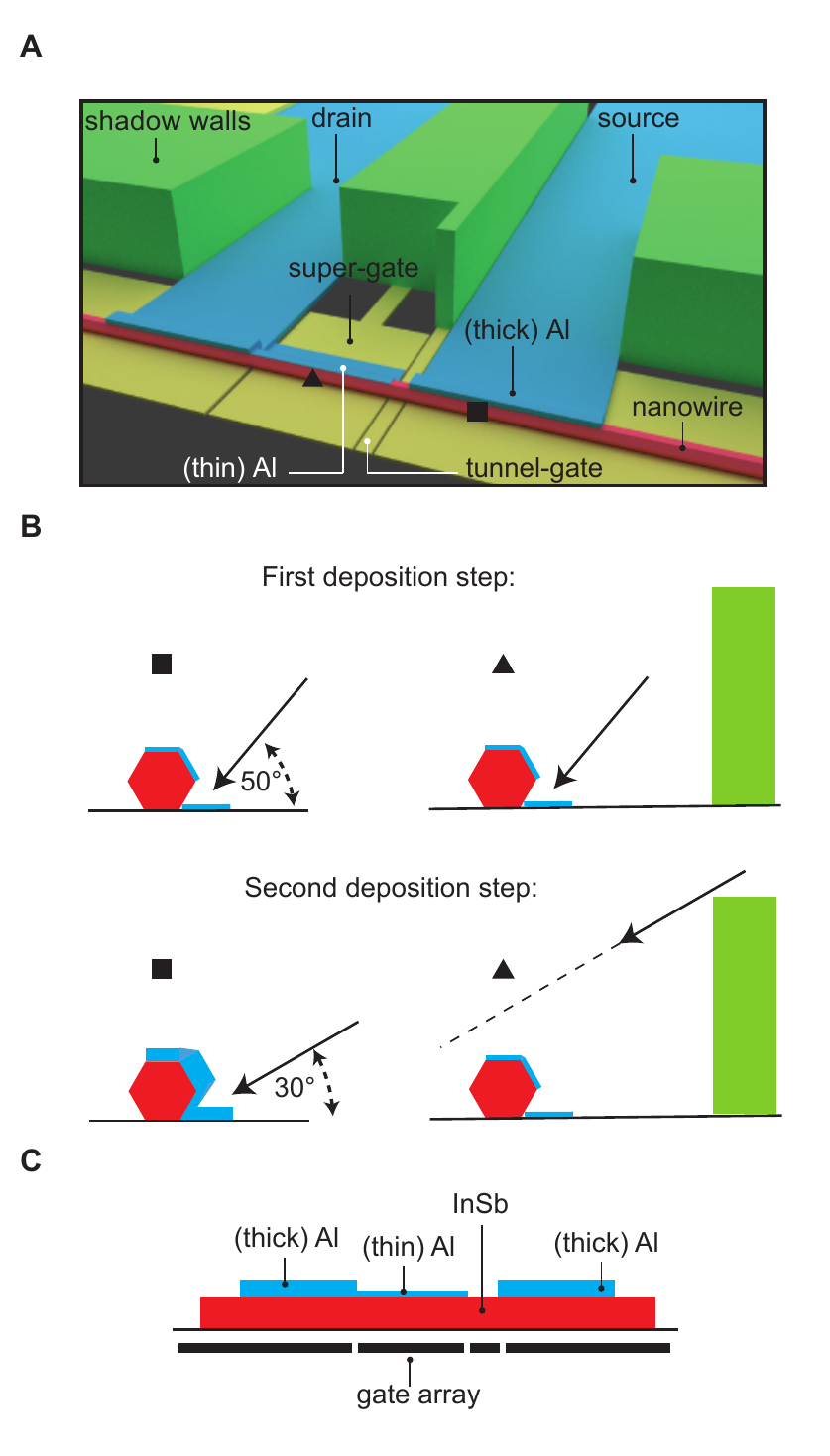}
	\caption{\textbf{Fabrication of asymmetric Josephson junctions.} (\textbf{A})~An illustration of an asymmetric Josephson junction device: the nanowire (in red) is separated from the bottom gates (in yellow) by a dielectric layer and is adjacent to shadow walls (in green). The superconducting film (in blue) covers the wire selectively. The relevant gates are the ‘super gate' and the ‘tunnel gate' whose actions are discussed later. (\textbf{B})~Cross-sections of the device taken at the two positions indicated by the square and triangle after the first (top panel) and the second evaporation step (bottom panel). (\textbf{C})~Longitudinal cross-section of the hybrid nanowire after the two steps.}
	\label{fig:Figure 1}
\end{figure}
Specifically, we exploit the properties of InSb nanowires and Al films, a combination of materials that is promising for the study of Majorana physics~\cite{Gazibegovic2017,deMoor2018,Heedt2020}. Nanowires are typically $\sim 10\,\upmu$m long~\cite{Badawy2019}, and the Al thin films can be grown homogeneously with thicknesses as low as $5\,$nm. 
The way the two materials are combined in-situ is illustrated in Fig.~\ref{fig:Figure 1}A. In brief, nanowires are transferred from the ‘growth’ to the ‘device’ substrate under an optical microscope. With accurate nano-manipulation, single nanowires (in red) are placed in the vicinity of dielectric shadow walls (in green) onto a gate oxide, which capacitively couples the wires to bottom gates (in yellow). 
Thanks to an atomic hydrogen cleaning step, the native oxide of the wires is removed without damaging the semiconductor crystal and introducing contaminations~\cite{Haworth2000,Tessler2006}. The superconductor is then deposited in-situ via e-gun evaporation at a substrate temperature of $\sim 140\,$K.\\
\indent Critically, the deposition is divided into two steps. First, we evaporate a thin Al layer at $50^{\circ}$ with respect to the substrate (top panel of Fig.~\ref{fig:Figure 1}B), and then a thick Al layer at $30^{\circ}$ (bottom panel of Fig.~\ref{fig:Figure 1}B). The two layers have a controlled thickness of $5-11\,$nm and $35-45\,$nm respectively (with $\sim 0.1\,$nm of accuracy). 
A longitudinal schematic cross-section of the device is illustrated in Fig.~\ref{fig:Figure 1}C to emphasize that, due to the shallow angle of the second evaporation and the position of the middle wall, only the two source and drain nanowire sections (Fig.~\ref{fig:Figure 1}A) are covered by the thick Al layer. 
This novel process enables the formation in a single shot of hybrid asymmetric Josephson junctions with Al leads of different thicknesses, and this controlled variation can be used as a knob to tune the superconducting properties of the junctions~\cite{Meservey1971}. 
When a normal metal (e.g., Pt) is deposited in the second step onto the thin Al layer, the source and drain nanowire sections act as normal leads due to the inverse proximity effect (see Fig.~S1).\\
\indent We emphasize that the ad-hoc design of shadow walls and the double-angle evaporation together with the isolation of the bond pads as discussed in ref.~\citenum{Heedt2020} allow for fabricating device without creating electrical shorts. Critically, the chip is mounted in a dilution refrigerator within a few hours after the evaporation, preventing the formation of chemical intermixing at the fragile InSb/Al interface.
The total elimination of detrimental processes such as heating steps, metal etching, lithography and electron microscopy makes our flow advantageous with respect to other state-of-the-art methods, and similar principles have been applied successfully in the context of carbon nanotube devices~\cite{Cao2005}.\\
\begin{figure*}[hbt]
	\centering
	\includegraphics[width=1\linewidth]{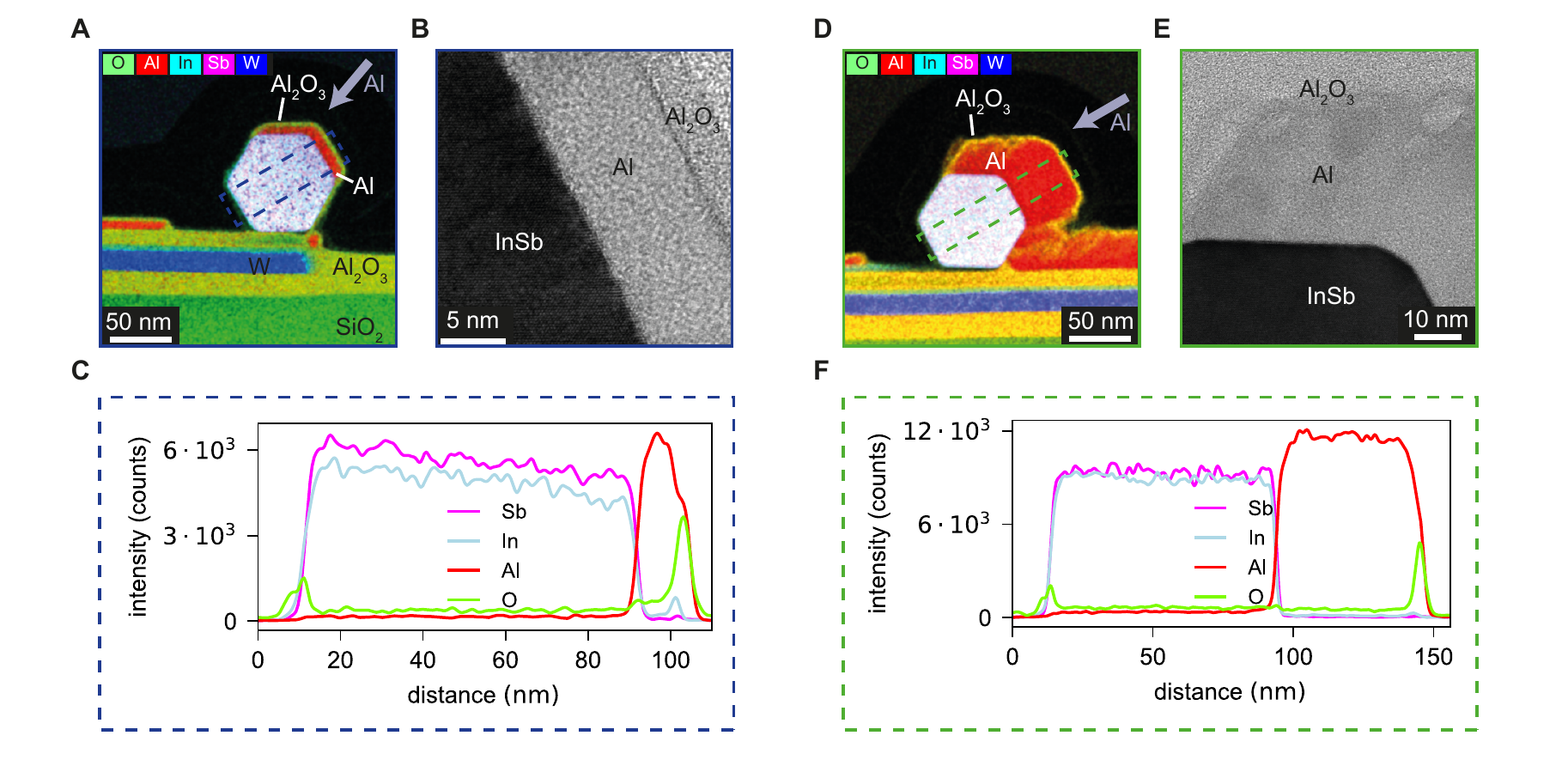}
	\caption{\textbf{Material analysis.} (\textbf{A})~Composite image of EDX elemental maps of the InSb nanowire covered with a thin layer of aluminum deposited at $50^{\circ}$ (cf. grey arrow). (\textbf{B})~Bright field (BF) STEM image of the InSb/Al interface. (\textbf{C})~Elemental line scan extracted from (\textbf{A}) perpendicular to the top-side facet of the nanowire. (\textbf{D})~Composite image of EDX elemental maps of the InSb nanowire covered with a thick layer of aluminum evaporated at $30^{\circ}$ (cf. grey arrow). (\textbf{E})~ADF STEM image focusing on the aluminum layer on the top facet of the nanowire. (\textbf{F})~Elemental line scan extracted from (\textbf{D}) perpendicular to the top-side facet of the nanowire.}
	\label{fig:Figure 2}
\end{figure*}
We evaluate the quality of the hybrid nanowires with transmission electron microscopy (TEM), and energy-dispersive x-ray analysis (EDX) performed on cross-sectional lamellas prepared via focused ion beam (FIB). The lamellas corresponding to Figs.~\ref{fig:Figure 2}A and D have been taken from cross-sections of nanowires with thin and thick Al coverage, respectively. Figs.~\ref{fig:Figure 2}B and E present typical bright-field scanning transmission electron microscopy images (BF STEM) of the thin- and thick-Al nanowire sections.\\
The EDX micrograph of Fig.~\ref{fig:Figure 2}A shows that, with the first deposition, the Al coating forms a continuous polycrystalline layer on two of the wire facets and on the substrate with a thickness of $6.5\,$nm on the top and $8.5\,$nm on the top-side facet. There is no connection between the thin Al shell on the wire and the thin Al on the substrate. Similarly, Fig.~\ref{fig:Figure 2}D illustrates that the thick-Al coverage is $26\,$nm and $49\,$nm depending on the facet. The wire exhibits three-facet coverage with a continuous metallic connection to the substrate, which allows for the creation of electrical contacts to the wire.\\ 
\indent Importantly, the interface between Al and InSb is sharp and clean, demonstrating a good connection between the two materials and no damage from hydrogen cleaning on the semiconductor. The EDX elemental mapping manifests an oxygen peak at the InSb/Al interface that is much weaker compared to the native oxide on the surface of the nanowire visible in the EDX line scan, highlighting the successful hydrogen cleaning treatment (Figs.~\ref{fig:Figure 2}C and F). \\
\begin{figure*}[hbt!]
	\includegraphics[width=1\linewidth]{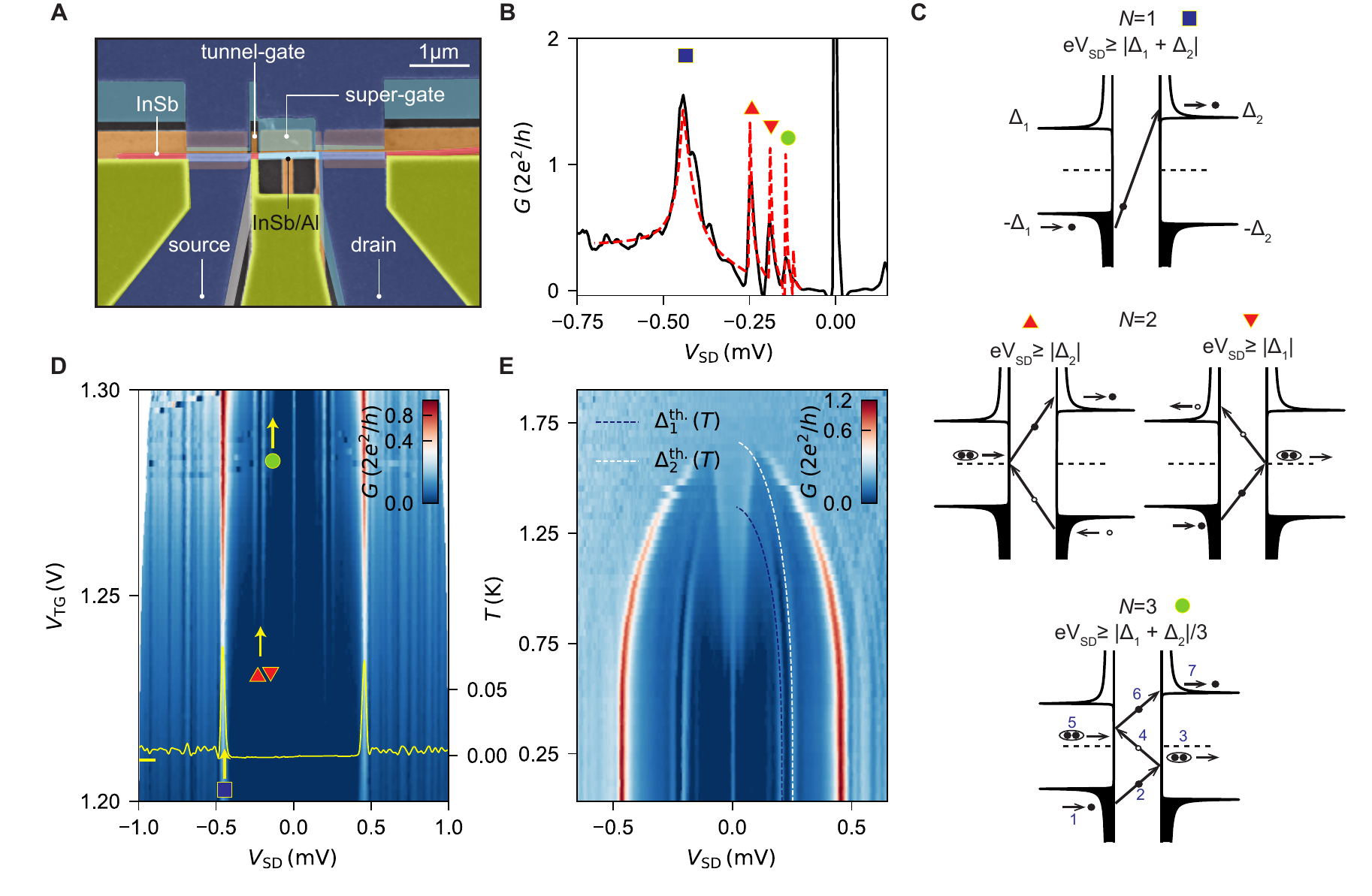}
	\caption{\textbf{Multiple Andreev reflections in asymmetric junctions.} (\textbf{A})~False-colour scanning electron micrograph of the first device. (\textbf{B})~$G$ vs.\ $V_{\mathrm{SD}}$ at $V_{\mathrm{TG}} = 1.38\,$V (black trace) and theoretical fit (dashed red trace). The blue square, red triangles and yellow circle indicate respectively multiple Andreev reflection peaks of the first, second and third order. (\textbf{C})~Schematic of the multiple Andreev reflections processes. Top, middle and bottom panels describe first, second and third orders, respectively. Electrons (holes) are shown as black (white) circles, and pairs of two electrons identify Cooper pairs. (\textbf{D})~Color map of $G$ vs.\ $V_{\mathrm{TG}}$ and $V_{\mathrm{SD}}$ displaying tunnelling conductance peaks at constant bias voltages. The square, the triangles and the circle correlate these peaks to the processes depicted in (\textbf{C}). The yellow trace is a line-cut at $V_{\mathrm{TG}} = 1.21\,$V, as indicated by the yellow tick. Values on the right y-axis are in units of $2e^2/h$. (\textbf{E})~Color map of $G$ vs. $V_{\mathrm{SD}}$ and temperature $T$ of the second device. Dashed lines indicate the expected BCS temperature dependence of the two gaps.}
	\label{fig:Figure 3}
\end{figure*}
\indent We validate our nanowire devices via low-temperature electrical transport.
We consider first an asymmetric Josephson junction device (scanning electron micrograph in Fig.~\ref{fig:Figure 3}A).
A DC bias voltage with a small AC excitation, $V_{\mathrm{SD}} + \delta V_{\mathrm{AC}}$, is applied between source and drain, yielding a current $I + \delta I_{\mathrm{AC}}$. Both the DC current and the differential conductance $G = \delta I_{\mathrm{AC}} / \delta V_{\mathrm{AC}}$ are measured in a dilution refrigerator with an electron temperature of $\sim 35\,$mK at its base temperature. 
The gate voltage $V_{\mathrm{TG}}$ applied at the ‘tunnel gate’ tunes the transmission of the junction, whereas the voltage $V_{\mathrm{SG}}$ at the ‘super gate’ controls the chemical potential of the proximitized wire.\\
The conductance through the device displays prominent peaks due to multiple Andreev reflections and a zero-bias peak due to Josephson supercurrent (Fig.~\ref{fig:Figure 3}B). Remarkably, these observations are found across all the measured devices, proving the strong and reproducible hybridization between the semi- and the superconductor (Fig.~S5). The observation of different orders of multiple Andreev reflections demonstrates that transport is phase-coherent across a length scale of multiple times the $100 \, \mathrm{nm}$-junction.\\
\indent In widely studied symmetric junctions, multiple Andreev reflection peaks arise at subharmonic values of the superconducting gap~\cite{Blonder1982,Octavio1983}. In asymmetric junctions, transport mechanisms such as the ones presented in Fig.~\ref{fig:Figure 3}C favour multiple Andreev processes at energies that relate to both gaps in a particular way. Odd orders manifest as conductance peaks at subharmonic values of the sum of the two gaps~\cite{Kuhlmann1994,Zimmermann1995}: 
\begin{equation}
V_{\mathrm{SD}} = \frac{\Delta_1 + \Delta_2}{Ne} \quad \mathrm{with} \, N=1, 3, 5,\ldots
\label{eq:odd_orders}
\end{equation}
where $\Delta_1$ and $\Delta_2$ are the small and the large gaps, respectively, $N$ is the order, and $e$ is the electronic charge. Differently, even orders give rise to doublets of peaks at subharmonic energies of both gaps $\Delta_i$ with $i = 1, 2$:
\begin{equation}
V_{\mathrm{SD}} = \frac{2 \Delta_i}{Ne} \quad \mathrm{with} \, N=2, 4, 6,\ldots
\label{eq:even_orders}
\end{equation}
\indent While the positions of the peaks depend on the magnitude of the two induced gaps, their intensity is related to the number and the transmission of the confined nanowire modes. To extract these parameters, we develop a theoretical model that accounts for different superconducting gaps in the two leads. 
In Fig.~\ref{fig:Figure 3}B we plot the result of the numerical calculation (red dashed trace) together with the experimental conductance (black trace). Here, the junction is found in the single-subband regime with transmission probability of $0.35$ and the gap values are $\Delta_1 = 192\,\upmu$eV, $\Delta_2 = 250\,\upmu$eV. The peaks at $eV_{\mathrm{SD}} = - (\Delta_1 + \Delta_2)$ (blue square), along with the doublet at $eV_{\mathrm{SD}} = - \Delta_{1,2}$ (red triangles) and the one at $eV_{\mathrm{SD}} = - (\Delta_1 + \Delta_2)/3$ (green circle), obey eqs.~\ref{eq:odd_orders} and \ref{eq:even_orders} for $N= 1, 2, 3$ perfectly. To the best of our knowledge, this is the first observation in nanowire-based devices of subharmonic structures where different gaps are involved, after the early investigations in planar Pb/InSb/Sn junctions~\cite{Kuhlmann1994,Zimmermann1995}. 
In Fig.~\ref{fig:Figure 3}D, we show the activation of these peaks upon varying $V_{\mathrm{TG}}$. While orders above $N=3$ are better resolved at higher junction transparency (Fig.~S2), the hard induced gap found in the tunnelling regime (see yellow trace) corroborates the high quality of the InSb/Al interface presented above.\\
\indent The difference in film thickness results also in two disparate superconducting critical temperatures: $T_{c1} \sim 1.35 \pm 0.05\,$K and $T_{c2} \sim 1.70 \pm 0.05\,$K, values that reflect the well-known enhancement in thin Al films with respect to the bulk value of $1.2\,$K~\cite{Cochran1958,Meservey1971}. We illustrate the difference of the two junction sides in Fig.~\ref{fig:Figure 3}E with the conductance map versus bias voltage and temperature taken on a second asymmetric junction device. The sub-harmonic conductance peaks with $N=2$, corresponding to $\pm \Delta_1$ and $\pm \Delta_2$, gradually shift to zero energy following well the BCS formula (blue and white dashed lines indicated by $\Delta_{1,2}^\mathrm{th.} \, (T)$). We note that the second device is conceptually similar to the first one, and exhibits comparable induced superconducting properties ($\Delta_1 = 203 \, \mathrm{\upmu eV} $ and $\Delta_2 = 253 \, \mathrm{\upmu eV}$ at base temperature). However, in the second device, the nanowire section coupled to the thin Al film is longer than the first ($1.5\,\upmu$m vs. $1.0\,\upmu$m). We emphasize that the ability to tune this parameter with ease (i.e., by shadow-wall design) is a novelty of our architecture, and it is relevant in topological circuits because it sets the maximum separation between emerging Majorana modes.\\
\begin{figure*}[hbt!]
	\includegraphics[width=1\linewidth]{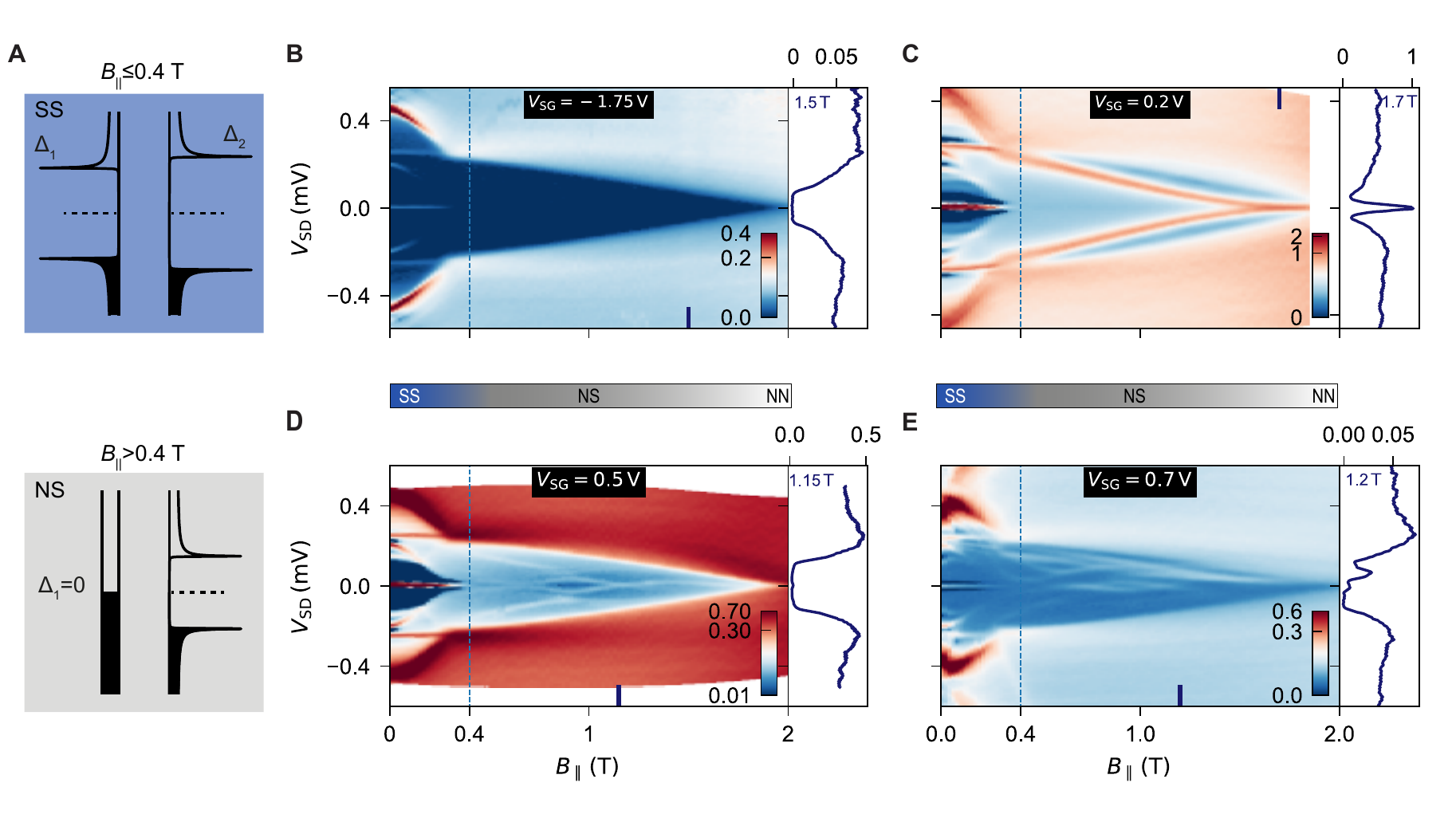}
	\caption{\textbf{Tunnelling spectroscopy in asymmetric Josephson junctions.} (\textbf{A})~ Top and bottom panels illustrate two schematics of the density of states at $B_\parallel \leq 0.4$ T, and $B_\parallel > 0.4$ in the asymmetric SS and NS junction regimes. The transition between the two regimes is marked with a vertical dashed line in the other panels. There, we display color maps of $G$ in units of $2e^2/h$ vs.\ $V_{\mathrm{SD}}$ and $B_{\parallel}$. The data sets differ in the value of $V_{\mathrm{SG}}$, which increases from (\textbf{B}) to (\textbf{E}): in (\textbf{B}) $V_{\mathrm{SG}}$ equals $-1.75\,$V, in (\textbf{C}) $0.2\,$V, in (\textbf{D}) $0.5\,$V and in (\textbf{E}) $0.7\,$V. Side panels show vertical line-cuts at the positions indicated by the blue lines and labels. The two horizontal bars illustrate the junction regimes as a function of magnetic field.}
	\label{fig:Figure 4}
\end{figure*}
The well controllable thickness of each shell deposited in our method allows for the creation of devices that can be tuned to realize different superconducting or normal elements. Such a transition is presented in Fig.~\ref{fig:Figure 4}, where we display results obtained on the second device. Upon increasing the magnetic field along the wire $(B_\parallel)$, the device transits from a Josephson junction (SS) into a normal-semiconductor-superconductor structure (NS), and eventually becomes a normal junction (NN) (Fig.~\ref{fig:Figure 4}A). The first transition is accompanied by the coalescence of the $\pm (\Delta_1 + \Delta_2)/e$ peaks into the $\pm \Delta_2/e$ peaks at $\sim 0.40\,$T, and the second occurs when $\Delta_2$ vanishes at $\sim 2 \, \mathrm{T}$. These two field boundaries allow for tunnelling-spectroscopy measurements in the NS configuration in a large magnetic field range relevant for topological superconductivity \cite{nijholt2016orbital}. A similar result was demonstrated in planar junctions on two-dimensional electron gases where the asymmetry of the critical fields was introduced by patterning one of the two leads much smaller than the superconducting coherence length \cite{Suominen2017,Nichele2017}.\\
\indent The versatility of our device preparation is also accompanied by the capability of tuning the device properties via the electric fields of the bottom gates.
By varying the super-gate voltage, for instance, it is possible to both change the number of bands in the wire and shift the electron density close to or far from the Al interface, renormalizing properties such as the hardness of the induced gap, the effective $g$-factor and the spin-orbit coupling~\cite{deMoor2018,antipov2018effects,winkler2019unified}.\\
At $V_{\mathrm{SG}} = -1.75\,$V, when the electron density is confined at the interface, bias spectroscopy reveals an Al-like hard-gap up to $\sim 1.8\,$T (Fig.~\ref{fig:Figure 4}B). When increasing super-gate voltage to $V_{\mathrm{SG}} = 0.2\,$V, we observe the first state (with an effective $g$-factor of $\sim 5.5$) coalescing with its own electron-hole symmetric partner in a zero-bias peak with conductance height of $\sim 2e^2/h$ (Fig.~\ref{fig:Figure 4}C). 
Although a $2e^2/h$-high peak is a hallmark of a resonant Andreev reflection into a Majorana state, the energy and peak height are tunable by both the super- and the tunnel-gates (Fig.~S7). This tunability suggests that this state might be a localized (i.e., non-topological) Andreev bound state located near the junction~\cite{Pan2020}.\\
\indent At $V_{\mathrm{SG}} = 0.5\,$V and $V_{\mathrm{SG}} = 0.7\,$V, when the positive gate voltage allows the occupation of a greater number of nanowire bands, we observe that, upon increasing the magnetic field, a low-energy state oscillates around zero energy (Figs.~\ref{fig:Figure 4}D and E). Additional subgap states shift down in energy (with an effective $g$-factor of $6-12$ in the first case, and $22$ at maximum in the second) and are repelled via the spin-orbit interaction, which in finite-length systems can couple states with different orbitals and spins~\cite{Stanescu2013,deMoor2018}. 
These states are tunable in energy by the super gate, but they are insensitive to variations in the tunnel gate (Fig.~S6). While this robustness in barrier transparency has been used in the past to substantiate the presence of Majorana modes, we stress that the population of multiple nanowire subbands (see simulations in Fig.~S9), together with orbital effects, result in a complex topological-phase diagram, making it arduous to assess the physical origin of the peaks in a normal-superconductor junction~\cite{nijholt2016orbital}.\\
\indent Because of this difficulty, future work will focus on three-terminal devices that enable probing both end-to-end subgap states correlations and, possibly, the presence of a topological gap~\cite{rosdahl2018andreev}. It is remarkable to note that the presented method, by allowing an accurate design and realization of standardized Majorana devices, will undoubtedly accelerate the progress in this direction.\\
One of the most significant challenges in the search for topological excitations in condensed matter is alleviating the complexity of the devices. This is a key aspect when it comes to reproducible measurements, high fabrication yield and eventually scalability. To this purpose, we introduce here an innovative technology to obtain hybrid nanowire junctions. By combining a unique double-angle evaporation with shadow walls, we demonstrate the possibility of completely eliminating the need for fabrication processing after the delicate semiconductor-superconductor interface is created. This method not only drastically reduces possible chemical contaminations and the deterioration of the interface but also results in reproducible and adjustable devices, with fast fabrication turnaround. Moreover, differently from previous shadowing methods without shadow walls~\cite{Gazibegovic2017,Carrad2019}, all the dimensions of the proximitized nanowire sections are accurately tunable.\\
While this study could not corroborate the presence of a topological state in hybrid nanowires, it lays the groundwork for future investigation of extremely long Majorana wires and for the fabrication of advanced devices such as hybrid two-path interferometers for the read-out of Majorana qubits~\cite{Fu2010,Whiticar2020,Borsoi2020}. 
Crucially, the versatility of the platform can spark rapid explorations of new material combinations toward a topological qubit, such as different semiconductors and superconductors, but also metals and ferromagnets. It also enables the direct synthesis of artificial Kitaev chains~\cite{Kitaev2001,Sau2012,Fulga2013}, facilitates advancement in Andreev qubits~\cite{Hays2018}, and the engineering of quasiparticle traps in devices such as superconducting qubits~\cite{Sun2012,Aumentado2004,vanVeen2018}.


\begin{acknowledgement}
The authors thank Mark Ammerlaan, Olaf Benningshof for valuable technical support and to Bernard van Heck and Giordano Scappucci for fruitful discussions. We also thank TNO for giving us access to their cleanroom tools. This work has been financially supported by the Dutch Organization for Scientific Research (NWO), the Foundation for Fundamental Research on Matter (FOM) and Microsoft Corporation Station Q. MPN acknowledges support within the POIR.04.04.00-00-3FD8/17 project as part of the HOMING programme of the Foundation for Polish Science co-financed by the European Union under the European Regional Development Fund. 
\end{acknowledgement}

\paragraph*{The authors declare no competing financial interests.}
\paragraph*{Data and materials availability}
The data that support the findings of this study are available at \url{https://doi.org/10.4121/uuid:9cdda62c-8bf1-4b48-8f0e-b9dceec546f7}.


\putbib
\end{bibunit}
\clearpage
\setcounter{equation}{0}
\setcounter{figure}{0}
\setcounter{table}{0}
\setcounter{page}{1}
\setcounter{section}{0}

\renewcommand{\thesection}{S\arabic{section}}  
\renewcommand{\thefigure}{S\arabic{figure}}
\renewcommand{\thetable}{S\arabic{table}} 
\renewcommand{\theequation}{S\arabic{equation}}
\renewcommand{\thefigure}{S\arabic{figure}}
\renewcommand{\bibnumfmt}[1]{[S#1]}
\renewcommand{\citenumfont}[1]{S#1}


\pagebreak

\titleformat*{\section}{\Large\bfseries}
\titleformat*{\subsection}{\large\bfseries}

\begin{bibunit}

\def\bibsection{\section*{Supplementary References}} 

\begin{center}
	\LARGE\textbf{Supplementary Information}\\
	\bigskip \bigskip \bigskip
	\LARGE{Single-shot fabrication of semiconducting-superconducting nanowire devices}\\
	\bigskip \bigskip \bigskip
	\LARGE{F. Borsoi \textit{et al.}}
\end{center}

\newpage
\section{Fabrication recipes}
\subsection*{Substrates fabrication}
Bottom gates are fabricated by reactive-ion etching a $\sim 17\,$nm thick W film with $\mathrm{SF_6}$ gas, and are then covered by a $\sim 18\,$nm layer of $\mathrm{Al_2 O_3}$ deposited via Atomic Layer Deposition. Shadow walls are created in the same top-down approach: a $\sim 700\,$nm thick layer of PECVD $\mathrm{Si_x N_y}$ is reactive-ion etched by $\mathrm{CHF_3}$. Shadow walls of the device presented in Fig.~\ref{fig:Figure Fabless} are made by patterning and developing a $\sim 1 \, \upmu \mathrm{m}$-thick layer of HSQ that is subsequently baked at $300 \, ^{\circ}$C. The substrates are then cleaned thoroughly for one hour with oxygen plasma to remove resist residues.

\subsection*{Nanowire growth and transfer}
Stemless InSb nanowires are grown with the method described in ref.~\cite{Badawy2019}. They are then deterministically transferred from the growth-substrate to the device-substrate and pushed in the vicinity of shadow walls using an optical microscope and a micro-manipulator. 

\subsection*{Semiconductor surface treatment and metal deposition}
\noindent After the nanowire transfer, the chip is loaded into the load-lock of the metal evaporator.
Here, the semiconductor oxide is gently removed with an atomic hydrogen cleaning treatment similar to ref.~\cite{Heedt2020}. 
A tungsten filament at $\sim 1700 \, ^{\circ}$C dissociates H$_2$ molecules into H$^*$ radicals which react with the oxygen at the nanowire surface and remove it. Typical parameters for this process are: a hydrogen flow of $2.2 \, \mathrm{ml/min}$, a process pressure of $6.3 \cdot 10^{-5} \, \mathrm{mbar}$ and a process time of $\sim 1 \,$hour. The holder onto which the chip is clamped is kept at $277 \, ^{\circ}$C for $\sim 3$ hours to ensure thermalization, and hydrogen cleaning is performed at this temperature.\\
The chip is then loaded from the load-lock into the main chamber of the evaporator. Here, it is cooled down and thermalized for an hour at $\sim 140\,$K. Thin Al is evaporated first at $50^{\circ}$ with respect to the substrate plane ($5-11\,$nm, measured by the evaporator crystal). Subsequently, Al (or Pt, depending on the device type) is deposited at $30^{\circ}$ ($35-45\,$nm, measured by the evaporator crystal). The deposition rate is maintained at $\sim 0.2 \, \mathrm{nm/min}$. The chip is brought back into the load-lock, where it is oxidized for 5 minutes in an oxygen pressure of $200\,$mTorr while still actively cooling the chip holder to maintain a temperature of $\sim 140\,$K. The load-lock is vented only when the chip has reached room temperature. 

\section{Transport measurements}
Electrical transport measurements were carried out in dilution refrigerators at a base temperature of approximately $15-20\,$mK and an electron temperature of approximately $35\,$mK. Lock-in conductance measurements were conducted at low frequency of $12-15\,$Hz.
The data presented in the main text was taken from two representative devices. Additional data of these two devices is shown in Figs.~\ref{fig:Multiple Andreev reflections},~\ref{fig:Crossover in the first device},~\ref{fig:Effect of magnetic field orientations},~\ref{fig:Subgap states} and ~\ref{fig:Zero-bias dependence}. 
In total, we fabricated and cooled down three chips with respectively 3, 4 and 6 nanowire devices that all manifested similar induced superconducting properties. Exemplary transport characteristics of additional asymmetric junction devices are shown in Fig.~\ref{fig:yield}, while data from a normal-superconducting junction is displayed in Fig.~\ref{fig:Figure Fabless}.

\section{Temperature dependence of the superconducting gap and determination of the effective \textit{g}-factor}
The temperature dependence of the superconducting gap displayed in Fig.~3E of the main text is expressed by the BCS theory as~\cite{Tinkham1996}:
\begin{equation}
\Delta^{\mathrm{th.}} (T) = \Delta_0 \cdot \mathrm{tanh} \left(1.74 \left( \frac{\Delta_0}{1.76 k_{\mathrm{B}} T} - 1 \right)^{\frac{1}{2}} \right)
\label{eq:temperature_dependence}
\end{equation}
where $\Delta_0$ is the gap at zero temperature and $k_{\mathrm{B}}$ the Boltzmann constant.
We estimate the effective $g$-factor of the subgap states in Fig.~4 by considering their average slope vs. the magnetic field as:
\begin{equation}
g = \frac{2 \Delta E}{\mu_{\mathrm{B}} \Delta B_{\parallel}} 
\label{eq:g-factor}
\end{equation}
where $ \Delta E $ is the variation in energy in a magnetic field range $\Delta B$, and $\mu_{\mathrm{B}}$ is the Bohr magneton. 

\section{TEM analysis}
TEM investigation was carried out at $200\,$keV with Thermo Fisher Talos transmission electron microscope equipped with Super-X EDX detector. TEM samples were prepared with focused-ion beam technique making use of Thermo Fisher Helios dual beam scanning electron microscope. Additional TEM analysis is shown in Fig.~\ref{fig:Addional material analysis}.

\section{Single-shot normal-superconductor junctions}
\begin{figure}[hbt!]
	\centering
	\includegraphics[width=0.5\linewidth]{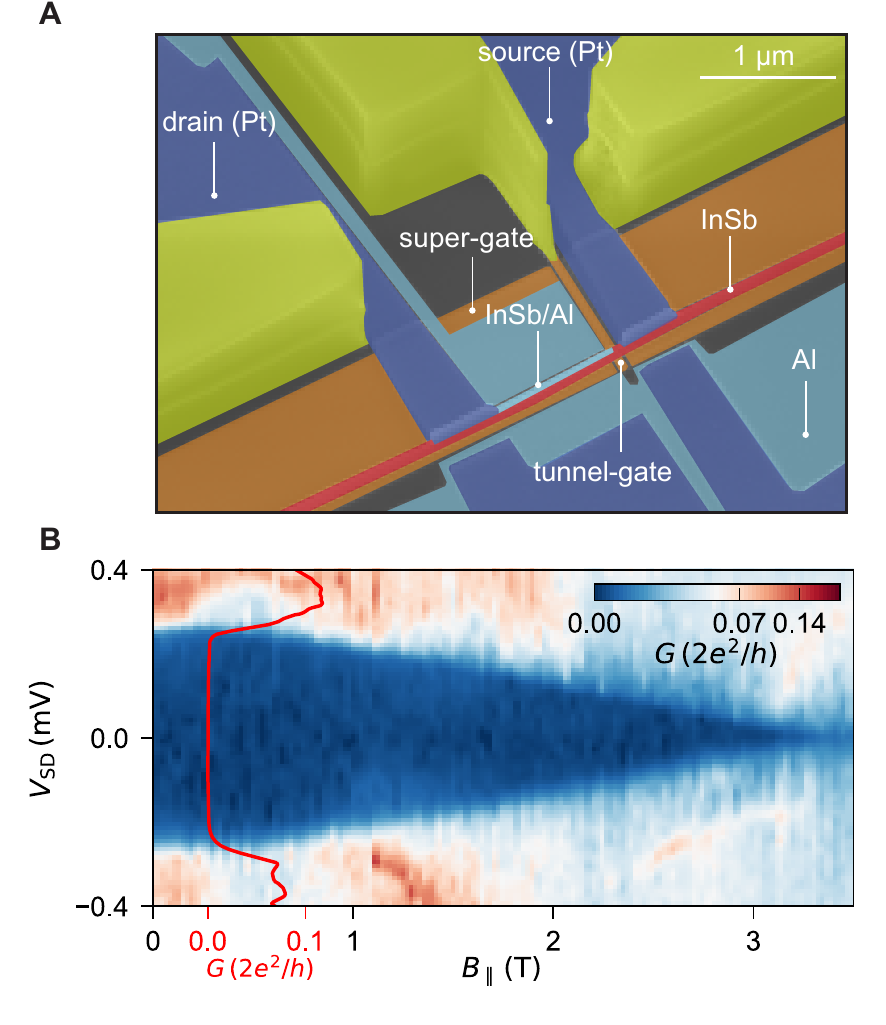}
	\caption{\textbf{Single-shot normal-superconductor junctions.} 
		(\textbf{A})~False-color scanning electron micrograph of a representative device based on Pt source and drain, an InSb nanowire, and Al as superconductor. (\textbf{B})~Bias spectroscopy as a function of parallel magnetic field at $V_{\mathrm{SG}} = 1\,$V. The red trace is taken at zero magnetic field. The noise in the measurement is due to the gate instability.}
	\label{fig:Figure Fabless}
\end{figure}

A second application of our technique is highlighted in the device presented in Fig.~\ref{fig:Figure Fabless}A. It represents a normal metal-superconductor junction realized in a very similar manner as the device in Fig.~3A of the main text. The substantial difference is that the leads are now made by a normal metal rather than a superconductor, enabling bias spectroscopy at zero magnetic field. For its realisation, after the deposition of a thin Al layer ($\sim 7\,$nm) at 50$^\circ$, we evaporate in-situ a film of Pt ($\sim 30\,$nm) at 30$^\circ$ that creates the two leads.\\
\indent Basic characterization of the device as a function of parallel magnetic field is shown in Fig.~\ref{fig:Figure Fabless}B. The measurement is taken with the super-gate voltage at $1\,$V, and with the junction strongly depleted by the tunnel gate. The red trace taken at zero field illustrates the quasiparticle density of states in the hybrid nanowire, which shows a hard induced gap of $\sim 275\,\upmu$eV. Upon increasing the parallel magnetic field, the in-gap conductance remains strongly suppressed up to $\sim 3\,$T, while the induced superconducting gap vanishes completely at $\sim 3.5\,$T (Fig.~\ref{fig:Figure Fabless}B). The increment of the critical field and of superconducting gap of this device with respect to the two asymmetric junction devices is due to the different Al thickness (here $7\,$nm versus $11.5$ and $12.5\,$nm of the first and second device respectively). We emphasize that the ability to form a nanowire-based junction in-situ involving different materials and with properties tunable by design has not been reported before. In the future, it is important to assess whether the combination of different metals induces strain in the nanowire system.

\section{Multiple Andreev reflections in the open regime}
\begin{figure*}[hbt!]
	\centering
	\includegraphics[width = 1 \linewidth]{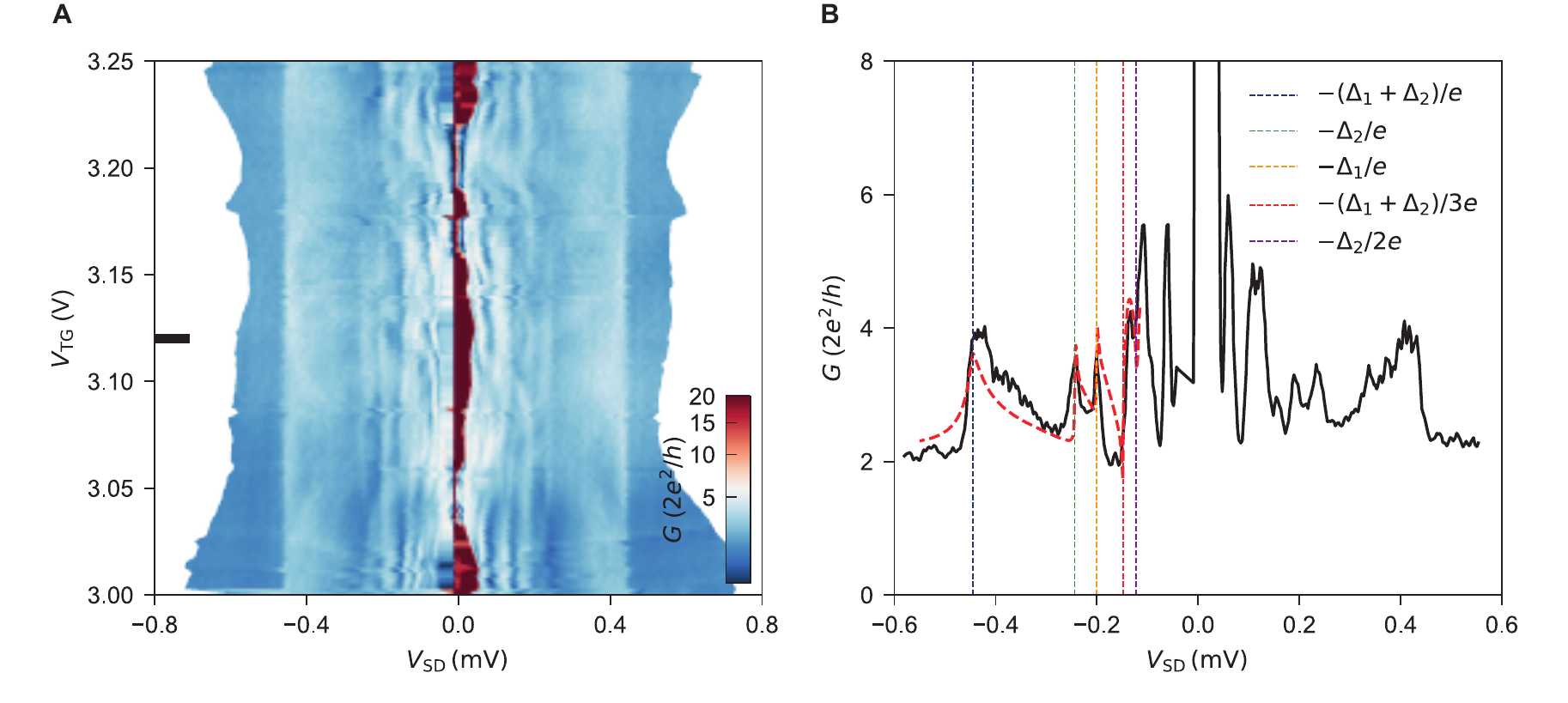}
	\caption{\textbf{Multiple Andreev Reflections in the open regime.} (\textbf{A})~$G$ vs $V_{\mathrm{SD}}$ and $V_{\mathrm{TG}}$ in the open regime. (\textbf{B})~The black trace is a line-cut of (\textbf{A}) taken at gate value indicated by the horizontal line, the red dashed trace is the best fit using the coherent-scattering model.}
	\label{fig:Multiple Andreev reflections}
\end{figure*}

Here, we display additional data related to the multiple Andreev processes occurring in asymmetric Al junctions. In Fig.~\ref{fig:Multiple Andreev reflections}A, we show a conductance map as a function of $V_{\mathrm{SD}}$ and $V_{\mathrm{TG}}$ in the open regime. Differently from the data-set presented in Fig.~3B and C in the main text where the out-of-gap conductance is $0.01-0.35 \cdot 2e^2/h$, here the out-of-gap conductance is $\sim 2 \cdot 2e^2/h$. In Fig.~\ref{fig:Multiple Andreev reflections}B, a conductance line-cut (black trace) is fitted with the coherent scattering model to evaluate the transmission though the one-dimensional channels (dashed red trace). Vertical line-cuts indicate the bias voltages at which the sub-harmonic features appear according to eqs.~1 and 2 of the main text. The fit parameters reveal the two induced gaps ($\Delta_1 = 201 \, \upmu eV$ and $\Delta_2 = 244 \, \upmu eV$) and indicate that three nanowire bands transmit with probabilities of 1, 0.82 and 0.33.

\section{Modelling multiple Andreev reflections in an asymmetric junction}
Below we introduce the model used to calculate conductance of a Josephson junction in which the two superconducting leads are characterized by different gap parameters: $\Delta_1$ and $\Delta_2$. In our model we consider a Josephson junction composed of two superconducting electrodes which are connected by a normal scattering region and are biased by $V_{\mathrm{SD}}$ voltage. 

We adopt the quasiparticle wave-function adjacent to the $L$'th lead in the form~\cite{Averin1995}:
\begin{equation}
\Psi_L = \sum_{n} \left[ \left(
\begin{array}{c}
A_{n}^L\\
B_{n}^L
\end{array} \right) e^{ikx}+
\left(\begin{array}{c}
C_{n}^L\\
D_{n}^L
\end{array} \right) e^{-ikx}
\right]e^{-i\left[E+neV_{\mathrm{SD}}\right]t/\hbar}.
\label{wf}
\end{equation}
$A_{n}^L$, $C_{n}^L$ correspond to the electron and $B_{n}^L$, $D_{n}^L$ to the hole amplitudes.

The scattering properties of the normal part of the junction are characterized by the scattering matrix:
\begin{equation}
S_{0}=\left(\begin{array}{cc}
r & t \\
t & -r \\
\end{array}\right),
\label{smat}
\end{equation}
with the amplitudes for the transmission $t=\sqrt{T}$ and reflection $r=\sqrt{1-T}$.

We assume that the junction is in short-junction regime and hence we take $S_0$ as  energy independent. We match the wave-functions Eq.~\eqref{wf} using $S_0$ obtaining the relations for electron and hole coefficients:
\begin{equation}
\left(\begin{array}{c}
A_{n}^\text{I}\\
A_{n+1}^{\text{II}}\\
\end{array}\right) = S_{0} \left(
\begin{array}{c}
C_{n}^\text{I}\\
C_{n+1}^\text{II}\\
\end{array} \right),
\label{sle1}
\end{equation}
and
\begin{equation}
\left(\begin{array}{c}
D_{n}^\text{I}\\
D_{n-1}^\text{II}\\
\end{array}\right) = S_{0} ^*\left(
\begin{array}{c}
B_{n}^\text{I}\\
B_{n-1}^\text{II}\\
\end{array} \right),
\label{slh1}
\end{equation}
respectively. The shifts of the indexes correspond to the changes of quasiparticle energies due to the bias voltage.

At each superconductor-normal interface we take into account Andreev reflections:
\begin{equation}
\begin{split}
\left(\begin{array}{c}
C_{n}^{L}\\
B_{n}^{L}
\end{array}\right)
=& \left(\begin{array}{c c}
a_{n}^L & 0\\
0 & a_{n}^L
\end{array} \right)
\left(
\begin{array}{c}
D_{n}^{L}\\
A_{n}^{L}
\end{array} \right)\\&+
\left( \begin{array}{c}
J_L(E+eV_L)\\
0
\end{array} \right)\frac{1}{\sqrt{2}}\delta_{p,e}\delta_{s,L}\kappa_{L}^+\\&+
\left(\begin{array}{c}
0\\
J_L(E-eV_L)
\end{array} \right)\frac{1}{\sqrt{2}}\delta_{p,h}\delta_{s,L}\kappa_{L}^-,
\end{split}
\label{AA_final}
\end{equation}
with the amplitude $a_{n}^L\equiv a^L(E+neV_{\mathrm{SD}})$, where,
\begin{small}
	\begin{equation}
	a^L(E) = \frac{1}{\Delta_L}\left\{
	\begin{array}{l l}
	E - \mathrm{sgn}(E)\sqrt{E^2-\Delta_L^2} & \quad |E|>\Delta_L\\
	E - i \sqrt{\Delta_L^2-E^2} & \quad |E|\leq \Delta_L
	\end{array} \right..
	\end{equation}
\end{small}

The last two terms in Eq.~(\ref{AA_final}) correspond to the source terms due to the quasiparticles incoming from the nearby superconducting electrodes~\cite{Nowak2019}, with the amplitude $J_L(E)=\sqrt{(1-a_L(E)^2)F_D(E)}$, where $F_D(E,T=30\;\mathrm{mK})$ is the Fermi distribution and where $\kappa_1^\pm = \delta_{n,0}$, $\kappa_2^\pm = \delta_{n,\pm1}$.

We calculate the DC current $I^L$ in the $L$'th lead as:
\begin{equation}
I^{L}=\frac{e}{\hbar\pi}\sum_{s=1,2}\sum_{p=e,h}\int_{-\infty}^{\infty}dE\sum_{n = -N_{\mathrm{max}}}^{N_\text{max}}(\mathbf{U}^{L*}_{n}\mathbf{U}^L_{n}
-\mathbf{V}^{L*}_{n}\mathbf{V}^L_{n}).
\label{CII}
\end{equation}
$\mathbf{U}^L_{n}=\left(A_{n}^L,B_{n}^L\right)^T$ and $\mathbf{V}^L_{n}=\left(C_{n}^L,D_{n}^L\right)^T$ are vectors that consist of the electron and hole amplitudes. The summation is carried over the position of the source term $s$ and $p$, the injected quasiparticle type. The DC current is subsequently used to calculate the conductance $G = dI^L/dV_{\mathrm{SD}}$.

\section{Fitting procedure}
To extract the induced gaps and the transmission probability of the normal part of the junctions studied in the experiment, we first calculate conductance of a multimode wire as a sum of contributions corresponding to $M$ modes of the transverse quantization as~\cite{Bardas1997}:
\begin{equation}
G_{\mathrm{theory}}(V_{\mathrm{SD}}) = \sum_{i}^{M}G_{i}(V_{\mathrm{SD}}, \Delta_{1}, \Delta_{2}, T_i).
\end{equation} 
$T_i$ determines the transmission probability for the $i$'th mode. Next, we fit the conductance obtained in theory to the experimental traces by minimizing $\chi = \int[G_\mathrm{{exp}}(V_{\mathrm{SD}})-G_{\mathrm{theory}}(V_{\mathrm{SD}})]^2 dV_{\mathrm{SD}}$ over $\Delta_{1}, \Delta_{2}, T_i$.
The comparison between the experimental conductance and the fitted theoretical traces are displayed in Fig.~3B in the main text (single-band regime) and in Fig.~\ref{fig:Multiple Andreev reflections}B (multi-band regime). In both scans, we see a single peak corresponding to the sum of the induced gaps for $eV_{\mathrm{SD}}=\pm |\Delta_1+\Delta_2|$. Higher-order subgap features are reflected by peaks at $eV_{\mathrm{SD}}=\pm \Delta_1$ and $eV_{\mathrm{SD}}=\pm \Delta_2$ and as a third-order process with the peak at $eV_{\mathrm{SD}}=\pm |(\Delta_1+\Delta_2)/3|$.

\section{Hard gap and magnetic field dependence in the first device}
\begin{figure*}[hbt!]
	\centering
	\includegraphics[width = 1 \linewidth]{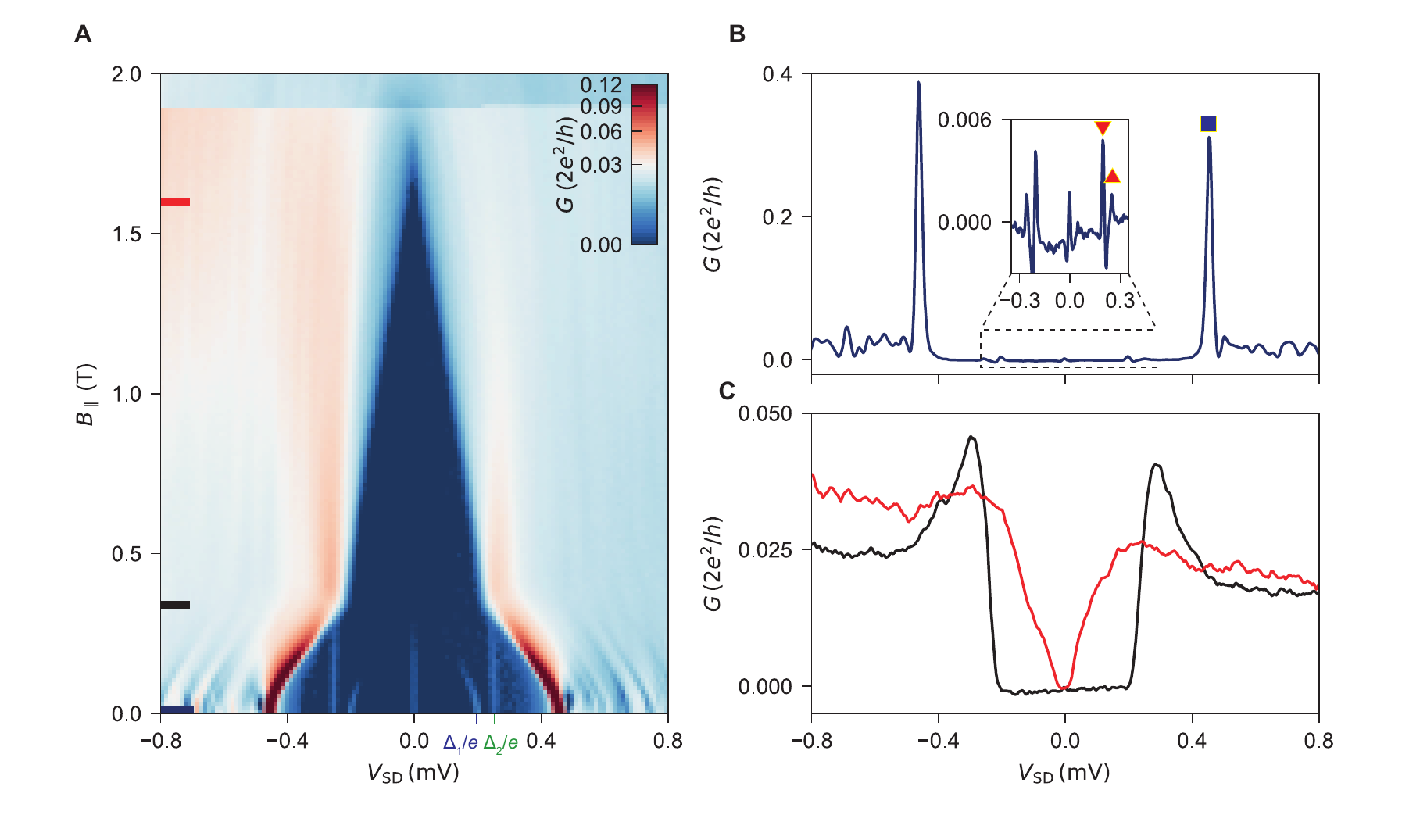}
	\caption{\textbf{Hard gap and magnetic field dependence in the first device.} (\textbf{A})~Color-map of $G$ vs. $V_{\mathrm{SD}}$ and $B_\parallel$. (\textbf{B}), (\textbf{C})~Line-cuts at the field values indicated by horizontal lines in (\textbf{A}). The inset in (\textbf{B}) shows a zoom-in of the multiple Andreev reflection peaks at bias voltage around zero. Peaks marked with the blue square and the red triangles correspond to the first and second orders, respectively.}
	\label{fig:Crossover in the first device}
\end{figure*}

In Fig.~\ref{fig:Crossover in the first device}A, we display a color map of $G$ vs. $V_{\mathrm{SD}}$ and $B_\parallel$ of the first device. Upon increasing magnetic field, the two-gap edges at symmetric bias values of $(\Delta_1 + \Delta_2)/e$ shift down in energy owing to the rapid dependence of $\Delta_1$ on the magnetic field (i.e., $\Delta_1$ is the superconducting gap of the thick-Al section). When $\Delta_1$ reaches zero at $\sim 0.34\,$T, the device transitions from a Josephson to a normal-superconductor junction. Tunnelling spectroscopy reveals a hard gap up to $\sim 1.6\,$T. Exemplary line-cuts are shown in Fig.~\ref{fig:Crossover in the first device}C, while the line-cut in Fig.~\ref{fig:Crossover in the first device}B at zero magnetic field manifests  Andreev reflections conductance peaks of the first order (blue square) and second order (red triangles). The small difference between the critical fields of the first and second device ($\sim 1.8\,$T vs. $\sim 2.0\,$T) is due to the slight difference between the two evaporation runs ($12.5\,$nm vs. $11.5\,$nm nominally at $50^\circ$ with respect to the substrate).

\section{Effect of magnetic field orientations}
\begin{figure*}[hbt!]
	\centering
	\includegraphics[width = 1 \linewidth]{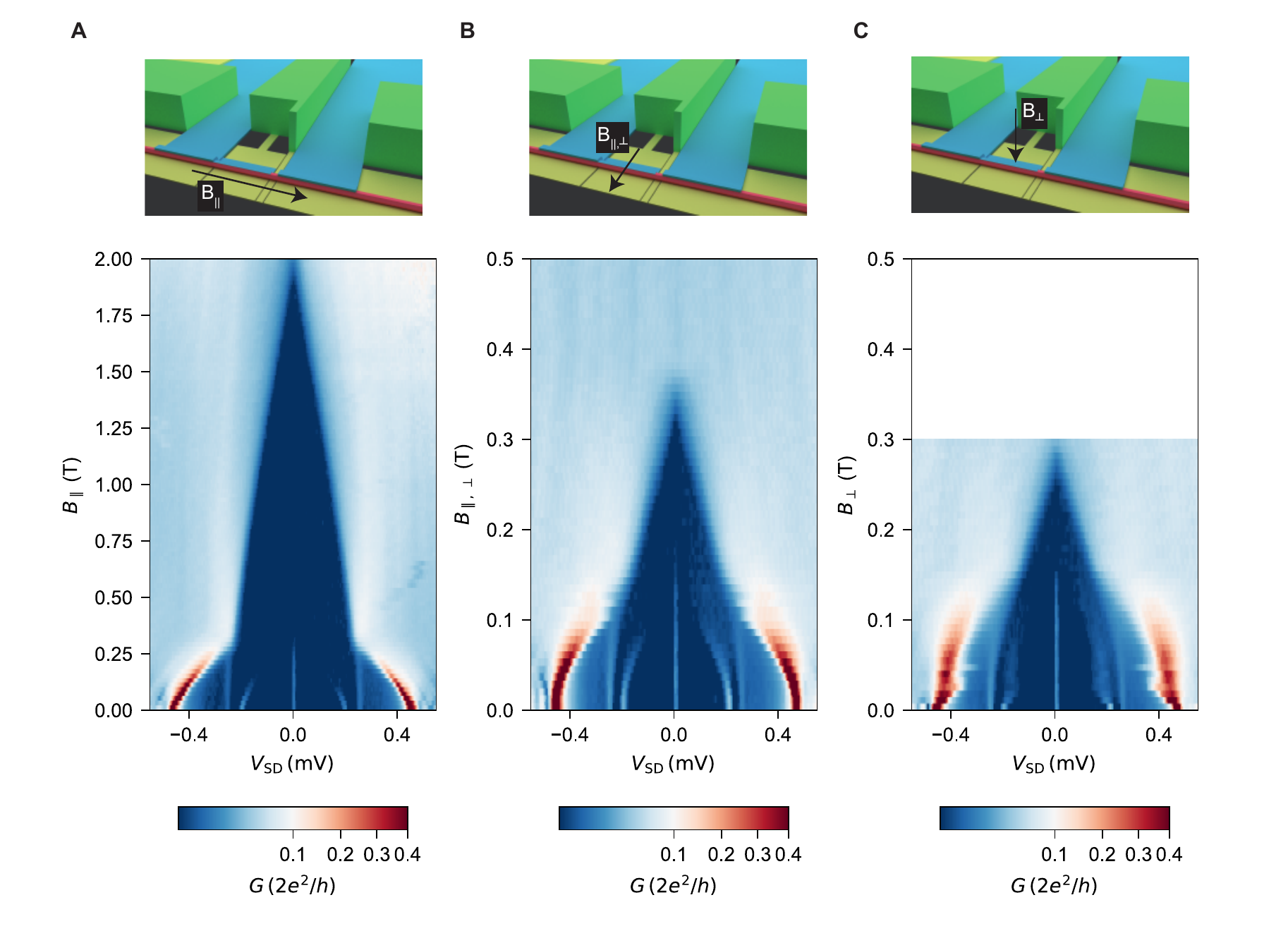}
	\caption{\textbf{Effect of magnetic field orientations.} Bias-spectroscopy measurements for three different magnetic field orientations. In (\textbf{A}) the field is parallel to the wire, in (\textbf{B}) is perpendicular to it and in-plane, in (\textbf{C}) it is orthogonal to the substrate.}
	\label{fig:Effect of magnetic field orientations}
\end{figure*}

We perform tunnelling-spectroscopy measurements in the second device for three different magnetic field orientations. The super-gate voltage is at $-1.75\,$V and the tunnel gate is at $1.3\,$V. In Fig.~\ref{fig:Effect of magnetic field orientations}A, the field $B_\parallel$ points along the wire direction, in panel B the field $B_{\parallel, \perp}$ is perpendicular to the wire and in-plane, and in panel C the field $B_\perp$ is oriented orthogonally to the substrate. The critical fields of the thin-Al hybrid nanowire are respectively $\sim 2\,$T, $\sim 0.37\,$T and $\sim 0.3\,$T. This observation is consistent with the geometry of the 2-facet Al shell.

\section{Zero-magnetic field transport in additional devices}
\begin{figure*}[hbt!]
	\centering
	\includegraphics[width = 1 \linewidth]{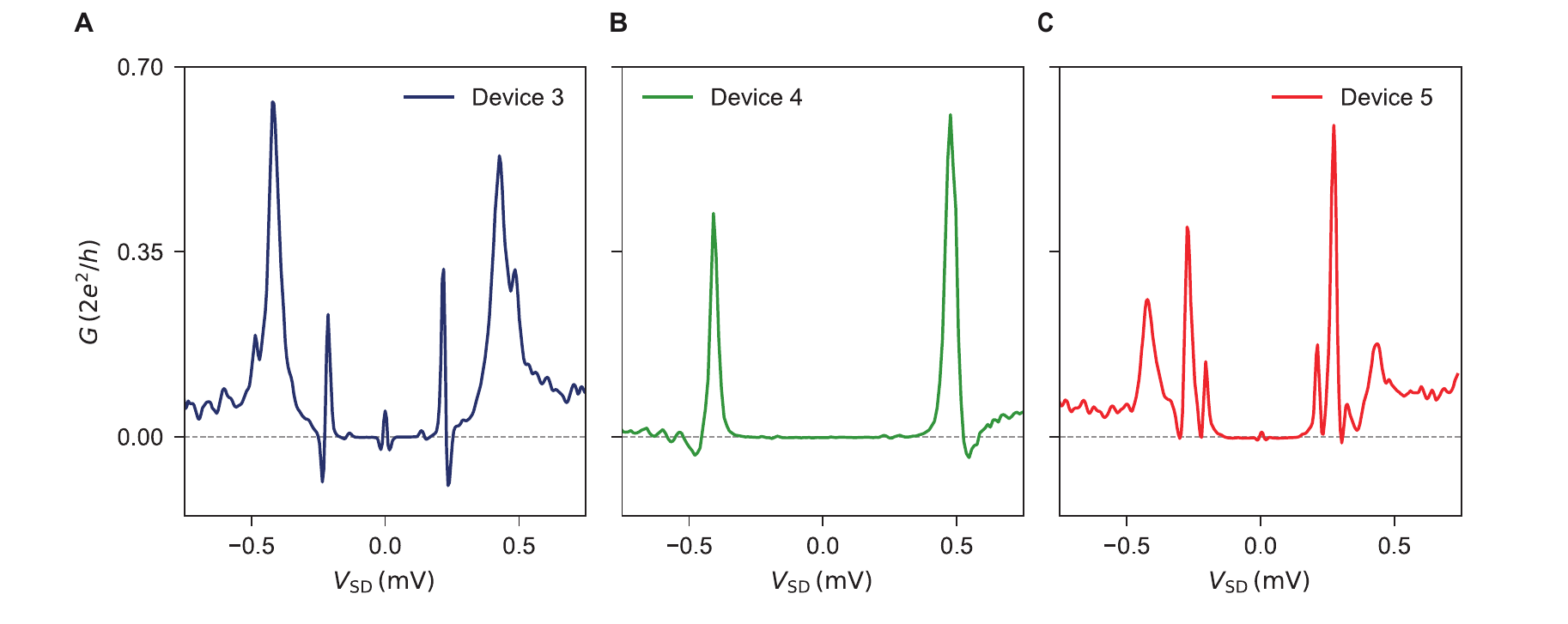}
	\caption{\textbf{Zero-magnetic field transport in additional devices.} $G$ versus $V_{\mathrm{SD}}$ for three additional devices. Devices 4 and 5 differ in evaporation run with respect to device 3. Each device has a different thin-Al section length, namely $2.0, 2.5$ and $1.0 \, \mathrm{\upmu m} $ for panels \textbf{A}, \textbf{B} and \textbf{C} respectively.}
	\label{fig:yield}
\end{figure*}

Here, we display basic transport characteristics of a third, a fourth and a fifth device respectively in Figs.~\ref{fig:yield}A, B, and C. In all three cases, the conductance in the tunnelling regime manifests strong suppression at low bias voltage and additional peaks due to multiple Andreev reflections. In Figs.~\ref{fig:yield}A, and C, the zero-bias supercurrent peak is also visible. These results, together with the data presented in the first and second devices, support the reproducibility of our fabrication process. 

\section{Subgap states: gate dependence}
\begin{figure*}[hbt!]
	\centering
	\includegraphics[width = 1 \linewidth]{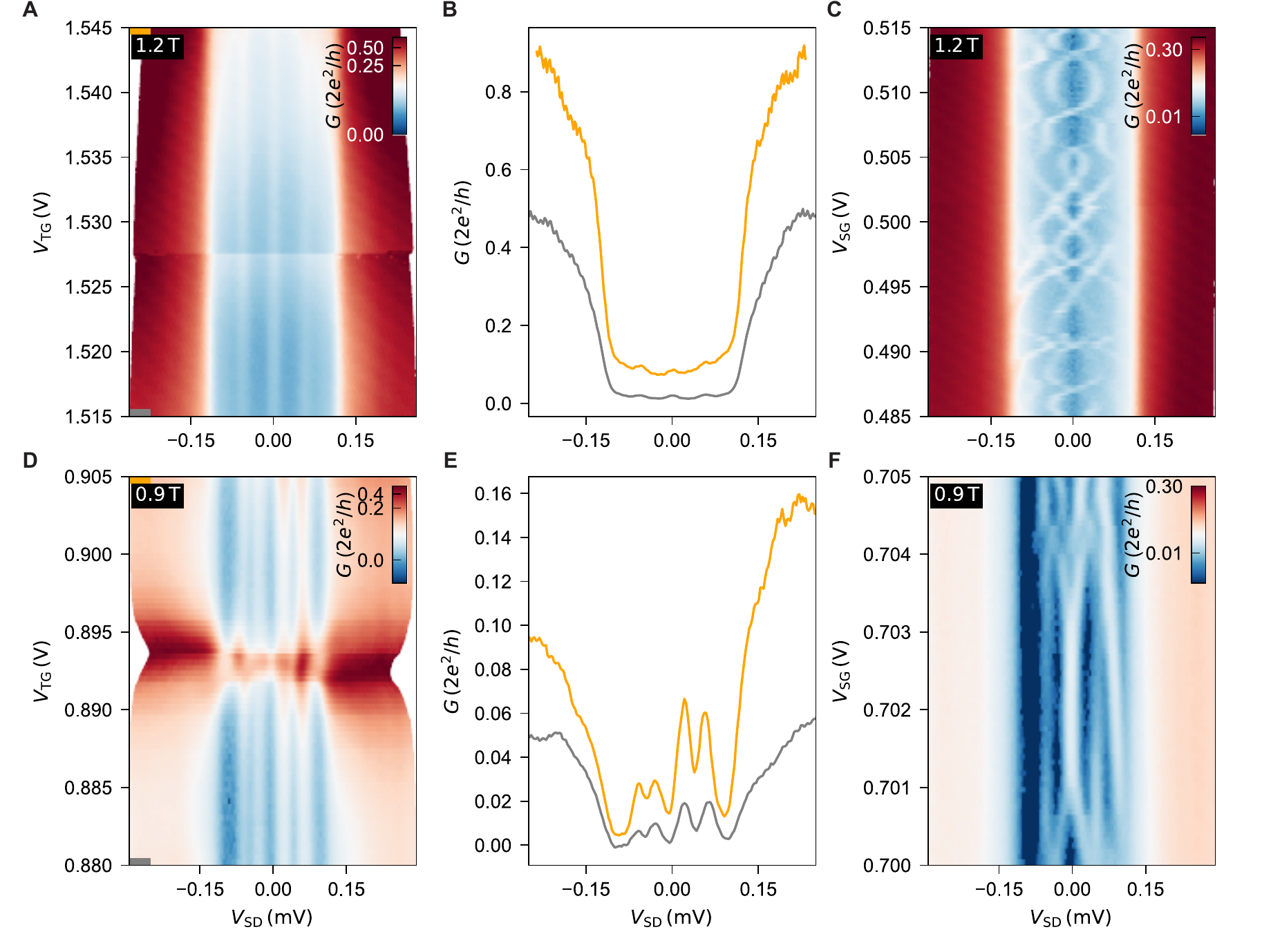}
	\caption{\textbf{Subgap states: gate dependence.} (\textbf{A}), (\textbf{B})~$G$ vs. $V_{\mathrm{TG}}$ and $V_{\mathrm{SD}}$ and line-cuts taken at the positions indicated by the horizontal lines at $B_{\parallel} = 1.2\,$T. (\textbf{C})~ $G$ vs. $V_{\mathrm{SG}}$ and $V_{\mathrm{SD}}$ at the same magnetic field. (\textbf{D}), (\textbf{E})~ $G$ vs. $V_{\mathrm{TG}}$ and $V_{\mathrm{SD}}$ and line-cuts at the positions indicated by the horizontal lines at $B_{\parallel} = 0.9\,$T. (\textbf{F})~$G$ vs. $V_{\mathrm{SG}}$ and $V_{\mathrm{SD}}$ at the same magnetic field.}
	\label{fig:Subgap states}
\end{figure*}

\begin{figure*}[hbt!]
	\centering
	\includegraphics[width = 1 \linewidth]{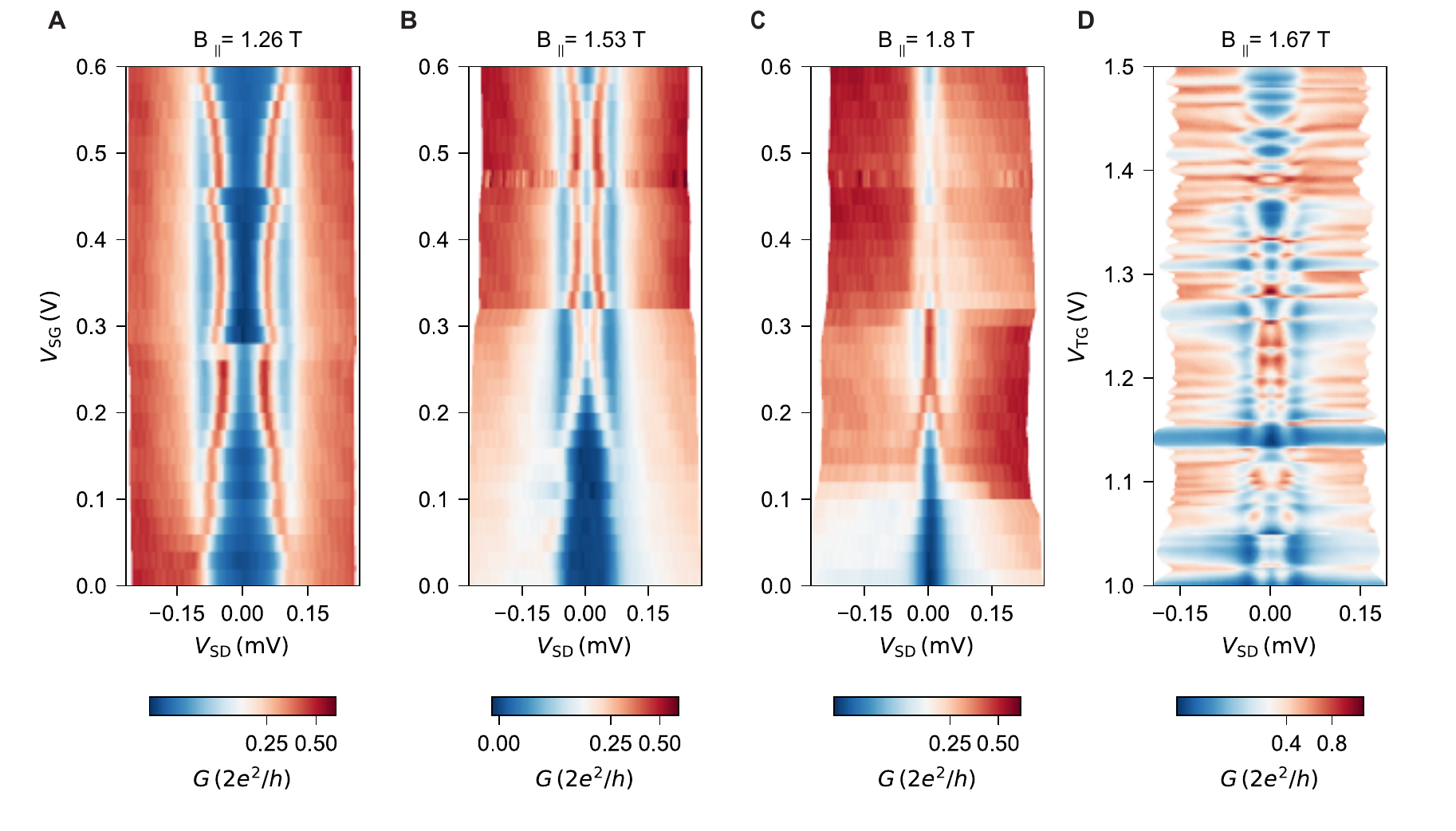}
	\caption{\textbf{Zero-bias peak gate dependence.} (\textbf{A}), (\textbf{B}) and (\textbf{C})~$G$ versus $V_{\mathrm{SD}}$ and $V_{\mathrm{SG}}$ taken with $B_{\parallel}$ at $1.26 \, \mathrm{T}$, $1.53 \, \mathrm{T}$ and $1.8 \, \mathrm{T}$, respectively. (\textbf{D})~$G$ versus $V_{\mathrm{SD}}$ and $V_{\mathrm{TG}}$ at $B_{\parallel} = 1.67\,$T.}
	\label{fig:Zero-bias dependence}
\end{figure*}

In Figs.~\ref{fig:Subgap states}A and C we show the tunnel- and super- gate dependences of the low-energy spectrum when $V_{\mathrm{SG}} = 0.5\,$V. The measurements are taken at $B_{\parallel} = 1.2\,$T, and enable to expand what is presented in Fig.~4D of the main text. The range of the two sweeps is similar ($30$ and $25 \, \mathrm{mV}$ respectively) to better illustrate the action of the two gates on the subgap states. The low-energy spectrum is not affected by the variation of the transmission of the junction, which changes by a factor of two with the voltage $V_{\mathrm{TG}}$ (see line-cuts in Fig.~\ref{fig:Subgap states}B). Differently, it does depend on the super gate along which oscillations take place, probably due to level repulsion between the subgap states. The same considerations are valid for the data-set shown in Figs.~\ref{fig:Subgap states}D, E and F taken for $V_{\mathrm{SG}} = 0.7\,$V at $B_{\parallel} = 0.9\,$T. The sudden increase in transmission in panel D is probably due to a resonance of an accidental quantum dot located at the junction.
In Fig.~\ref{fig:Zero-bias dependence}, we show the super-gate and tunnel-gate dependence of the subgap state presented in Fig.~4C of the main text for different magnetic field. The large tunability of the peak suggests that this state might be a localized (i.e., non-topological) Andreev bound state located near the junction~\cite{Pan2020}.

\section{Additional material analysis}
\begin{figure*}[hbt!]
	\centering
	\includegraphics[width = 1 \linewidth]{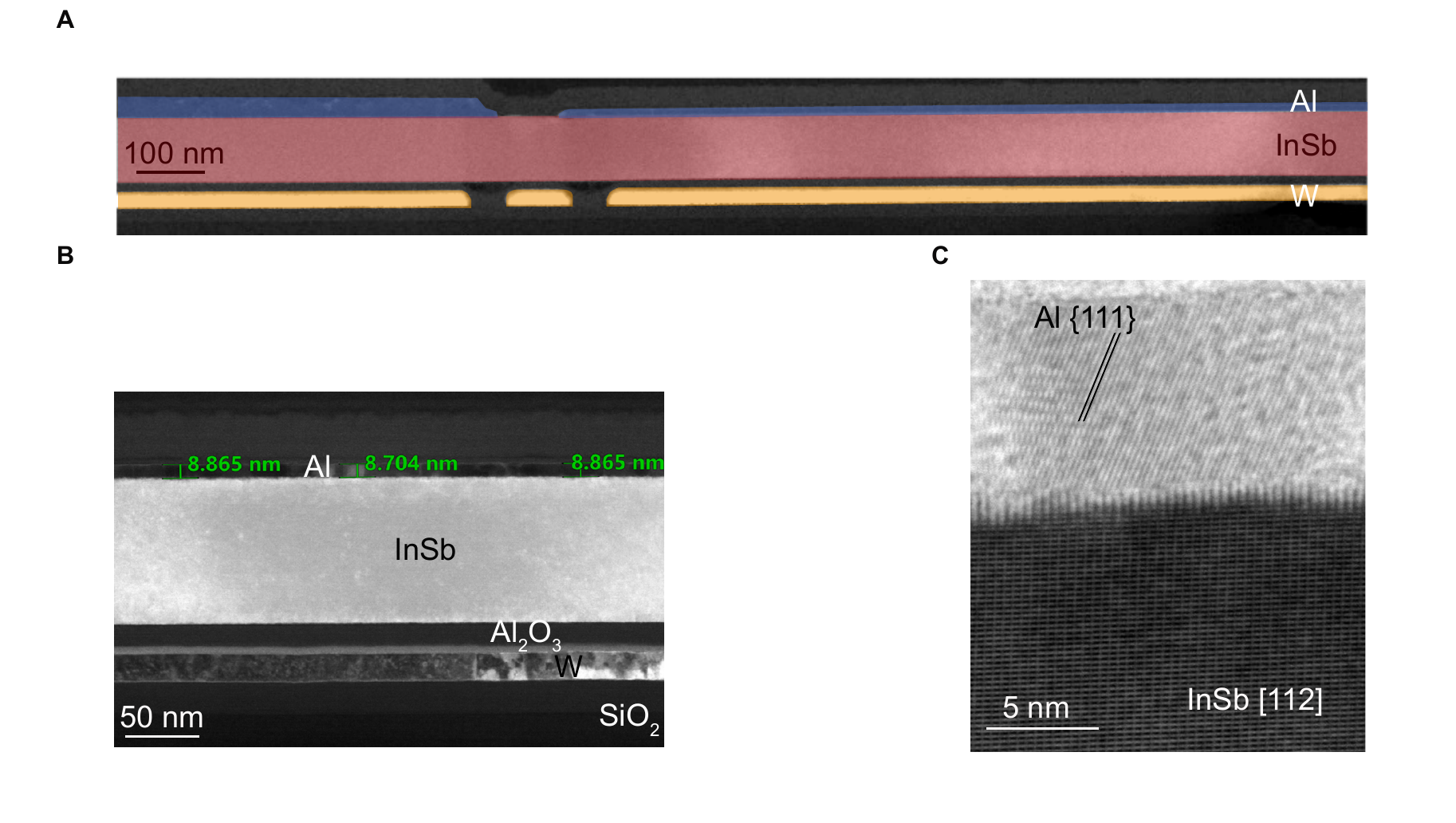}
	\caption{\textbf{Additional material analysis.} (\textbf{A})~ False-colour HAADF STEM image along the longitudinal FIB cut of the InSb nanowire showing the junction with the thin Al layer on the right size on top of the wire and the thick one on the left side. The nanowire is shown in red, the gates in orange and the Al in blue. (\textbf{B})~ Magnified ADF STEM image of the InSb nanowire with the thin Al coating. (\textbf{C})~ High-resolution ABF STEM image of the InSb/Al interface in the region with thin Al coating. }
	\label{fig:Addional material analysis}
\end{figure*}

TEM investigation was carried out at $200\,$keV with Thermo Fisher Talos transmission electron microscope equipped with Super-X EDX detector. TEM samples were prepared with focused-ion beam technique making use of Thermo Fisher Helios dual beam scanning electron microscope. Extensive STEM imaging with high-angle annular dark field (HAADF), annular dark field (ADF) and bright field (BF) detectors has not revealed any extended crystallographic defects in the wire, such as stacking faults or dislocations (Figs.~\ref{fig:Addional material analysis}A and B as examples). The thickness of Al clearly changes at the junction region from thicker layer on the left side of the junction to the thinner one on the right (Fig.~\ref{fig:Addional material analysis}A). High-resolution annular bright field (ABF) STEM image shows clean InSb/Al interface and good crystallinity of both materials. The roughness of the interface in Fig.~\ref{fig:Addional material analysis}C is attributed to its tilt in the viewing direction caused by the difficulty in the sample preparation of the hexagonal nanowire along the main axis.

\section{Simulations of proximitized wire}
\begin{figure*}[hbt!]
	\centering
	\includegraphics[width = 1 \linewidth]{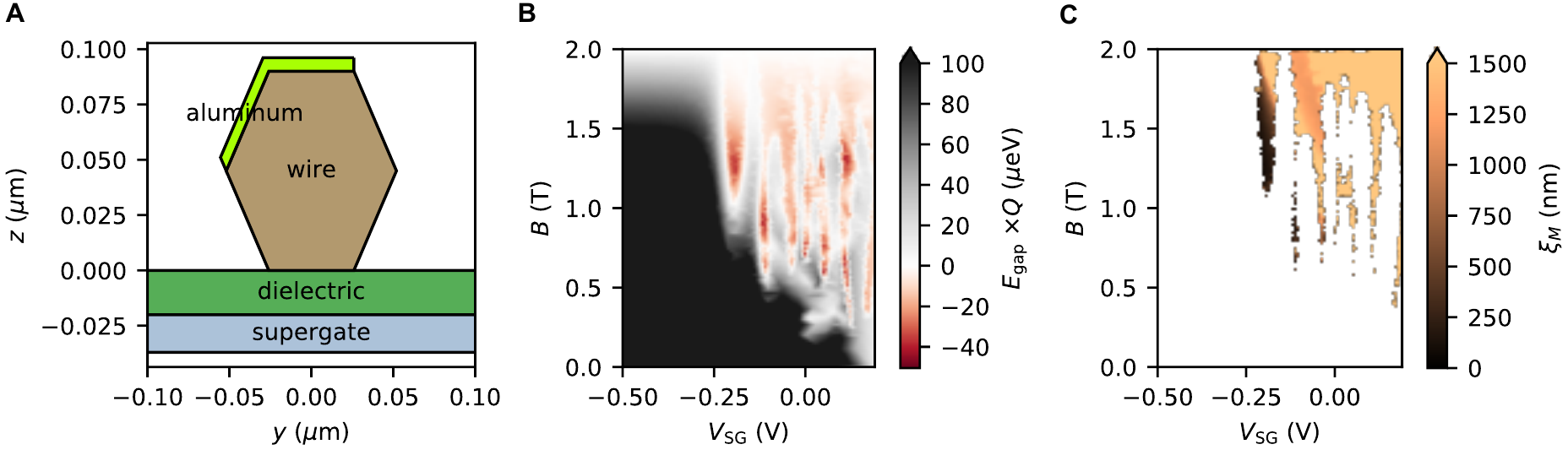}
	\caption{\textbf{Phase-diagram simulation of topological gap and Majorana length.} (\textbf{A})~ Geometry used in the simulation; the wire is infinite in the $x$ direction. (\textbf{B})~ Product of gap and topological invariant; positive indicates a trivial phase and negative a topological phase. (\textbf{C})~ Majorana length.}
	\label{fig:sim1}
\end{figure*}

\begin{figure*}[hbt!]
	\centering
	\includegraphics[width = 1 \linewidth]{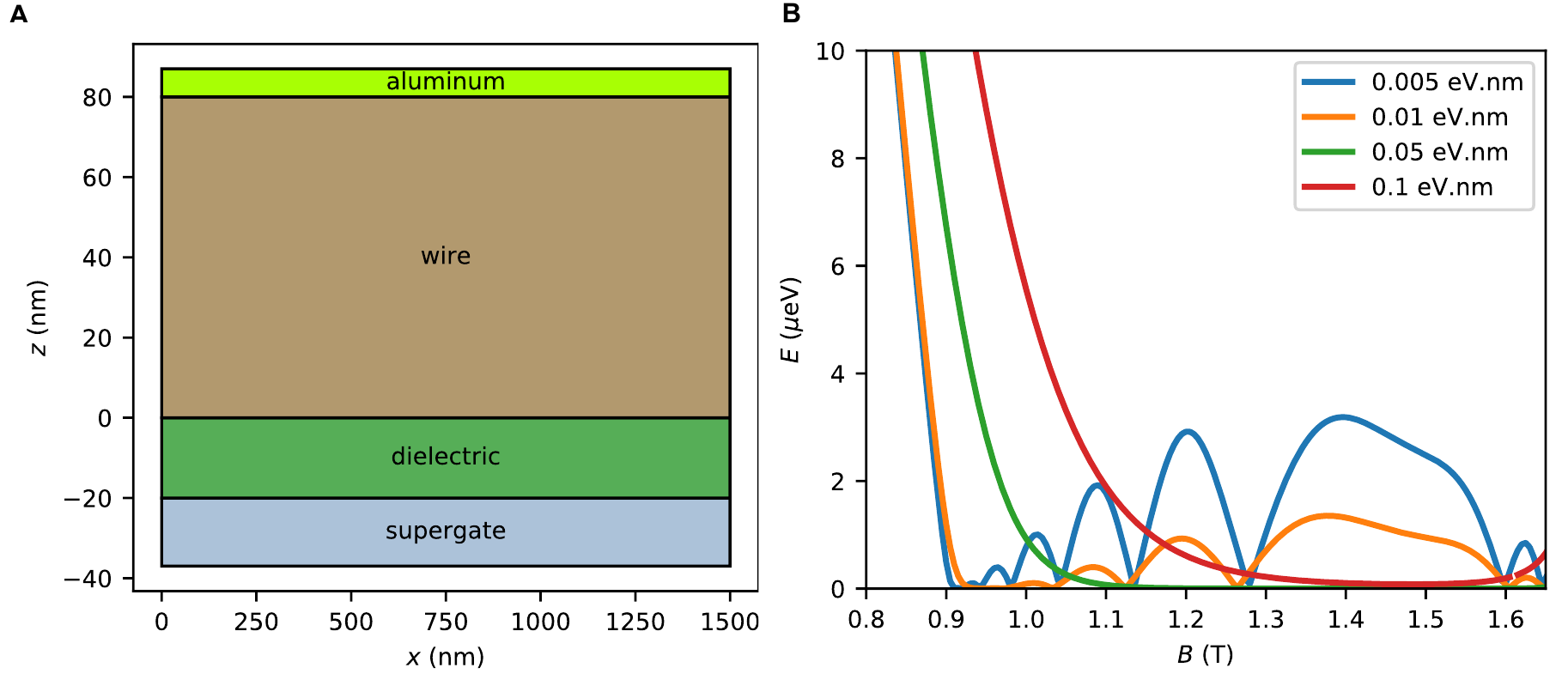}
	\caption{\textbf{2D simulation of finite nanowire.} (\textbf{A}) Geometry used in simulation. (\textbf{B}) Dependence of the energy of the lowest state in a magnetic field. The curves are the results of different simulations with varying $\alpha_R$.}
	\label{fig:sim2}
\end{figure*}

In Fig.~4 of the main text, we illustrate the behaviour of the spectrum of the wire as a function of magnetic field and super-gate voltage. The latter influences the number of subgap states and their effective $g$-factor, and consequently it would effect the onset of a potential transition to the topological regime. 
In order to study this phenomenon and the oscillatory states observed in Fig.~4D and E of the main text, we perform two types of simulations. 
The first one is phase-diagram calculation of the topological gap as a function of super-gate potential and magnetic field. We proceed along the line of~ref.~\citenum{vaitiekenas2020flux}: the electrostatic potential is solved within the Thomas-Fermi approximation, which is used in a Bogoliubov-de Gennes equation discretized on a finite-difference grid. The superconductor properties are integrated into self-energy boundary conditions, as discussed in~\cite{vaitiekenas2020flux}, which allow to calculate the gap ($E_{\rm gap}$) and topological invariant ($Q$). We include orbital effects of the magnetic field through Peierl's substitution, and the closing of the superconducting gap as:
\begin{equation}
\Delta(B) = \Delta_0 \sqrt{1-\frac{B^2}{B_c^2}} 
\label{eq:gap_closing}
\end{equation} 
with $B_c = 2$ T, $\Delta_0 = 0.25$ meV. The band offset between Al and InSb is set at 50 meV, and the Rashba SOC is $\alpha_R = 0.1$ eV nm.
We also compute $\xi_M$, the length of the Majorana wave-function, from the eigenvalue decomposition of the translation operator (see ref.~\cite{nijholt2016orbital}).
The geometry of our model is shown in Fig.~\ref{fig:sim1}A. 
In Fig.~\ref{fig:sim1}B, the product of the gap and the topological invariant as a function of magnetic field and supergate $V_{\rm SG}$ is plotted. We see the appearance of topological phases starting at $V_{\rm SG} ~\simeq -0.25$ V, and a maximum topological gap of 50 $\upmu$eV. We emphasize that this simulation cannot predict the exact position of the topological phases in parameter space, because the band offset and the density of interface charges at the interface are unknown. 
In Fig.~\ref{fig:sim1}C, the associated Majorana length is shown for the same parameters: the smallest length of 150 nm is observed in the first band, but it quickly increases with magnetic field to up to more than $1\, \mathrm{\upmu m}$. Given that the length of Al in the experimental device is 1.5 $\upmu$m, it is expected that the two Majorana wave-functions should be strongly hybridized.
The second type of simulations performed aims to make a more detailed analysis of the oscillations of the lowest energy state when entering a regime of hybridized Majoranas. To that end, we apply the same methodology as in the previous simulation, but in the cross-section of a finite wire in the XZ plane. We then solve for the states with energies below the gap in this 2D system. The geometry is shown in Fig~\ref{fig:sim2}A.
It is important to highlight here that this simulation reproduces the coupling to the superconductor and the electrostatic profile less accurately than the previous one. Because the superconductor appears only on the top, there is less hybridization in this case, and the Rashba spin-orbit coupling is renormalized less strongly. Therefore, in order to see Majorana lengths comparable to the one in the first band of Fig.~\ref{fig:sim1}C we need to reduce effectively the Rashba coefficient (at least by a factor 10). We use again $B_c = 2$ T, $\Delta_0 = 0.25$ meV, and a band offset of 50 meV, but this time we sweep over several values of $\alpha_R$. The results are shown in Fig.~\ref{fig:sim2}B for the first topological phase, at $V_{\rm SG} = 0.016$ V. \\
For the two values of expected Rashba coefficient in InSb wires (0.05 and 0.1 eV$\cdot$nm) we do not find any oscillation. For a Rashba coefficient ten times smaller, oscillations appear, with an amplitude of a few $\upmu$eV increasing with magnetic field. This is much smaller than the amplitude of the measured oscillations. 
From these simulations we can conclude that without any disorder or inhomogeneity in the system, it is unlikely that states in the topological regime show oscillations decaying with magnetic field.

\newpage

\putbib
\end{bibunit}


\begin{thebibliography}{10}

\bibitem{Alicea2012}
J.~Alicea.
\newblock New directions in the pursuit of \mbox{M}ajorana fermions in solid
  state systems.
\newblock {\em Rep. Prog. Phys.}, 75(7):076501, jun 2012.

\bibitem{antipov2018effects}
A.~E. Antipov, A.~Bargerbos, G.~W. Winkler, B.~Bauer, E.~Rossi, and R.~M.
  Lutchyn.
\newblock Effects of gate-induced electric fields on semiconductor
  \mbox{M}ajorana nanowires.
\newblock {\em Phys. Rev. X}, 8(3):031041, 2018.

\bibitem{Aumentado2004}
J.~Aumentado, Mark~W. Keller, John~M. Martinis, and M.~H. Devoret.
\newblock Nonequilibrium quasiparticles and $2e$ periodicity in
  single-\mbox{C}ooper-pair transistors.
\newblock {\em Phys. Rev. Lett.}, 92:066802, Feb 2004.

\bibitem{Badawy2019}
G.~Badawy, S.~Gazibegovic, F.~Borsoi, S.~Heedt, C.-A. Wang, S.~Koelling, M.~A.
  Verheijen, L.~P. Kouwenhoven, and E.~P. A.~M. Bakkers.
\newblock High mobility stemless \mbox{InSb} nanowires.
\newblock {\em Nano Lett.}, 19(6):3575--3582, 2019.

\bibitem{Bjergfeld2019}
M.~Bjergfelt, D.~J. Carrad, T.~Kanne, M.~Aagesen, E.~M. Fiordaliso, E.~Johnson,
  B.~Shojaei, C.~J. Palmstr{\o}m, P.~Krogstrup, T.~S. Jespersen, and
  J.~Nyg{\aa}rd.
\newblock Superconducting vanadium/indium-arsenide hybrid nanowires.
\newblock {\em Nanotechnology}, 30(29):294005, May 2019.

\bibitem{Blonder1982}
G.~E. Blonder, M.~Tinkham, and T.~M. Klapwijk.
\newblock Transition from metallic to tunneling regimes in superconducting
  microconstrictions: excess current, charge imbalance, and supercurrent
  conversion.
\newblock {\em Phys. Rev. B}, 25:4515--4532, Apr 1982.

\bibitem{Borsoi2020}
F.~Borsoi, K.~Zuo, S.~Gazibegovic, R.~L. M.~Op het Veld, E.~P. A.~M. Bakkers,
  L.~P. Kouwenhoven, and S.~Heedt.
\newblock Transmission phase read-out of a large quantum dot in a nanowire
  interferometer.
\newblock {\em Nat. Commun.}, 11:3666, 2020.

\bibitem{Boscherini1987}
F.~Boscherini, Y.~Shapira, C.~Capasso, C.~Aldao, M.~del Giudice, and J.~H.
  Weaver.
\newblock Exchange reaction, clustering, and surface segregation at the
  \mbox{Al/InSb}(110) interface.
\newblock {\em Phys. Rev. B}, 35:9580--9585, Jun 1987.

\bibitem{Cao2005}
J.~Cao, Q.~Wang, and H.~Dai.
\newblock Electron transport in very clean, as-grown suspended carbon
  nanotubes.
\newblock {\em Nat. Mater.}, 4:745--749, 2005.

\bibitem{Carrad2019}
D.~J. Carrad, M.~Bjergfelt, T.~Kanne, M.~Aagesen, F.~Krizek, E.~M. Fiordaliso,
  E.~Johnson, J.~Nyg{\aa}rd, and T.~S. Jespersen.
\newblock Shadow epitaxy for in situ growth of generic
  semiconductor/superconductor hybrids.
\newblock {\em Adv. Mat.}, 32(23):1908411, 2020.

\bibitem{Cochran1958}
J.~F. Cochran and D.~E. Mapother.
\newblock Superconducting transition in aluminum.
\newblock {\em Phys. Rev.}, 111:132--142, Jul 1958.

\bibitem{deMoor2019}
M.~W.~A de~Moor.
\newblock {\em Quantum transport in nanowire networks}.
\newblock PhD thesis, Delft University of Technology, Apr 2019.

\bibitem{deMoor2018}
M.~W.~A. de~Moor, J.~D.~S. Bommer, D.~Xu, G.~W. Winkler, A.~E. Antipov,
  A.~Bargerbos, G.~Wang, N.~van Loo, R.~L. M.~Op het Veld, S.~Gazibegovic,
  D.~Car, J.~A. Logan, M.~Pendharkar, J.~S. Lee, E.~P. A.~M. Bakkers, C.~J.
  Palmstr{\o}m, R.~M. Lutchyn, L.~P. Kouwenhoven, and H.~Zhang.
\newblock Electric field tunable superconductor-semiconductor coupling in
  majorana nanowires.
\newblock {\em New J. of Phys.}, 20(10):103049, oct 2018.

\bibitem{Fu2010}
L.~Fu.
\newblock Electron teleportation via \mbox{M}ajorana bound states in a
  mesoscopic superconductor.
\newblock {\em Phys. Rev. Lett.}, 104:056402, Feb 2010.

\bibitem{Fulga2013}
I.~C. Fulga, A.~Haim, A.~R. Akhmerov, and Y.~Oreg.
\newblock Adaptive tuning of \mbox{M}ajorana fermions in a quantum dot chain.
\newblock {\em New J. Phys.}, 15(4):045020, Apr 2013.

\bibitem{Gazibegovic2019}
S.~Gazibegovic.
\newblock {\em Bottom-up grown \mbox{InSb} nanowire quantum devices}.
\newblock PhD thesis, \mbox{Technische} Universiteit Eindhoven, May 2019.

\bibitem{Gazibegovic2017}
S.~Gazibegovic, D.~Car, H.~Zhang, S.~C. Balk, J.~A. Logan, M.~W.~A. de~Moor,
  M.~C. Cassidy, R.~Schmits, D.~Xu, G.~Wang, P.~Krogstrup, R.~L.~M. \mbox{Op
  het Veld}, K.~Zuo, Y.~Vos, J.~Shen, D.~Bouman, B.~Shojaei, D.~Pennachio,
  J.~S. Lee, P.~J. van Veldhoven, S.~Koelling, M.~A. Verheijen, L.~P.
  Kouwenhoven, C.~J. Palmstr{\o}m, and E.~P. A.~M. Bakkers.
\newblock Epitaxy of advanced nanowire quantum devices.
\newblock {\em Nature}, 584:434--438, 2017.

\bibitem{Haworth2000}
L.~Haworth, J.~Lu, D.~I. Westwood, and J.~E. MacDonald.
\newblock Atomic hydrogen cleaning, nitriding and annealing \mbox{InSb} (100).
\newblock {\em Appl. Surf. Sci.}, 166(1):253 -- 258, 2000.

\bibitem{Hays2018}
M.~Hays, G.~de~Lange, K.~Serniak, D.~J. van Woerkom, D.~Bouman, P.~Krogstrup,
  J.~Nyg\aa{}rd, A.~Geresdi, and M.~H. Devoret.
\newblock Direct microwave measurement of andreev-bound-state dynamics in a
  semiconductor-nanowire josephson junction.
\newblock {\em Phys. Rev. Lett.}, 121:047001, Jul 2018.

\bibitem{Heedt2020}
S.~Heedt, M.~Quintero-P\'erez, F.~Borsoi, A.~Fursina, N.~van Loo, G.~P. Mazur,
  M.~P. Nowak, M.~Ammerlaan, K.~Li, S.~Korneychuk, J.~Shen, M.~A.~Y. van~de
  Poll, G.~Badawy, S.~Gazibegovic, K.~van Hoogdalem, E.~P. A.~M. Bakkers, and
  L.~P. Kouwenhoven.
\newblock Shadow-wall lithography of ballistic superconductor-semiconductor
  quantum devices.
\newblock {\em ArXiv e-prints}, 2007.14383, 2020.

\bibitem{Hyart2013}
T.~Hyart, B.~van Heck, I.~C. Fulga, M.~Burrello, A.~R. Akhmerov, and C.~W.~J.
  Beenakker.
\newblock Flux-controlled quantum computation with \mbox{M}ajorana fermions.
\newblock {\em Phys. Rev. B}, 88:035121, Jul 2013.

\bibitem{Karzig2017}
T.~Karzig, C.~Knapp, R.~M. Lutchyn, P.~Bonderson, M.~B. Hastings, C.~Nayak,
  J.~Alicea, K.~Flensberg, S.~Plugge, Y.~Oreg, C.~M. Marcus, and M.~H.
  Freedman.
\newblock Scalable designs for quasiparticle-poisoning-protected topological
  quantum computation with \mbox{M}ajorana zero modes.
\newblock {\em Phys. Rev. B}, 95:235305, 2017.

\bibitem{Khan2020}
Sabbir~A. Khan, Charalampos Lampadaris, Ajuan Cui, Lukas Stampfer, Yu~Liu,
  Sebastian~J. Pauka, Martin~E. Cachaza, Elisabetta~M. Fiordaliso, Jung-Hyun
  Kang, Svetlana Korneychuk, Timo Mutas, Joachim~E. Sestoft, Filip Krizek, Rawa
  Tanta, Maja~C. Cassidy, Thomas~S. Jespersen, and Peter Krogstrup.
\newblock Highly transparent gatable superconducting shadow junctions.
\newblock {\em ACS Nano}, May 2020.

\bibitem{Kitaev2001}
A.~Yu. Kitaev.
\newblock Unpaired \mbox{M}ajorana fermions in quantum wires.
\newblock {\em Phys.-Uspekhi}, 44(10S):131--136, 2001.

\bibitem{Krogstrup2015}
P.~Krogstrup, N.~L.~B. Ziino, W.~Chang, S.~M. Albrecht, M.~H. Madsen,
  E.~Johnson, J.~Nyg{\aa}rd, C.~M. Marcus, and T.~S. Jespersen.
\newblock Epitaxy of semiconductor-superconductor nanowires.
\newblock {\em Nat. Mater.}, 14:400, 2015.

\bibitem{Kuhlmann1994}
M.~Kuhlmann, U.~Zimmermann, D.~Dikin, S.~Abens, K.~Keck, and V.~M. Dmitriev.
\newblock Andreev-reflection in semiconductor-coupled superconducting
  weak-links.
\newblock {\em Z. Phys. B}, 96(1):13--24, Mar 1994.

\bibitem{Lutchyn2010}
R.~M. Lutchyn, J.~D. Sau, and S.~Das~Sarma.
\newblock Majorana fermions and a topological phase transition in
  semiconductor-superconductor heterostructures.
\newblock {\em Phys. Rev. Lett.}, 105:077001, 2010.

\bibitem{Meservey1971}
R.~Meservey and P.~M. Tedrow.
\newblock Properties of very thin aluminum films.
\newblock {\em J. Appl. Phys.}, 42(1):51--53, 1971.

\bibitem{Nichele2017}
F.~Nichele, A.~C.~C. Drachmann, A.~M. Whiticar, E.~C.~T. O'Farrell, H.~J.
  Suominen, A.~Fornieri, T.~Wang, G.~C. Gardner, C.~Thomas, A.~T. Hatke,
  P.~Krogstrup, M.~J. Manfra, K.~Flensberg, and C.~M. Marcus.
\newblock Scaling of \mbox{M}ajorana zero-bias conductance peaks.
\newblock {\em Phys. Rev. Lett.}, 119:136803, 2017.

\bibitem{nijholt2016orbital}
B.~Nijholt and A.~R. Akhmerov.
\newblock Orbital effect of magnetic field on the \mbox{M}ajorana phase
  diagram.
\newblock {\em Phys. Rev. B}, 93(23):235434, 2016.

\bibitem{Octavio1983}
M.~Octavio, M.~Tinkham, G.~E. Blonder, and T.~M. Klapwijk.
\newblock Subharmonic energy-gap structure in superconducting constrictions.
\newblock {\em Phys. Rev. B}, 27:6739--6746, 1983.

\bibitem{Oreg2010}
Y.~Oreg, G.~Refael, and F.~von Oppen.
\newblock Helical liquids and \mbox{M}ajorana bound states in quantum wires.
\newblock {\em Phys. Rev. Lett.}, 105:177002, 2010.

\bibitem{Pan2020}
H.~Pan, W.~S. Cole, J.~D. Sau, and S.~Das~Sarma.
\newblock Generic quantized zero-bias conductance peaks in
  superconductor-semiconductor hybrid structures.
\newblock {\em Phys. Rev. B}, 101:024506, Jan 2020.

\bibitem{Pei1988}
J.~H. Pei, R.~Manzke, and C.~G. Olson.
\newblock Exchange reaction at the \mbox{Al-InSb} (110) interface.
\newblock {\em Chin. J. of Phys.}, 26, Jun 1988.

\bibitem{Plugge2017}
S.~Plugge, A.~Rasmussen, R.~Egger, and K.~Flensberg.
\newblock Majorana box qubits.
\newblock {\em New J. Phys.}, 19(1):012001, 2017.

\bibitem{rosdahl2018andreev}
T.~Rosdahl, A.~Vuik, M.~Kjaergaard, and A.~R. Akhmerov.
\newblock Andreev rectifier: a nonlocal conductance signature of topological
  phase transitions.
\newblock {\em Phys. Rev. B}, 97(4):045421, 2018.

\bibitem{Sau2012}
Jay~D. Sau and S.~Das Sarma.
\newblock Realizing a robust practical majorana chain in a
  quantum-dot-superconductor linear array.
\newblock {\em Nat. Comm.}, 3(1):964, 2012.

\bibitem{Sporken1988}
R.~Sporken, P.~Xhonneux, R.~Caudano, and J.P. Delrue.
\newblock The formation of the \mbox{Al-InSb}(110) interface.
\newblock {\em Surf. Sci.}, 193(1):47 -- 56, 1988.

\bibitem{Stanescu2013}
Tudor~D. Stanescu, Roman~M. Lutchyn, and S.~Das~Sarma.
\newblock Dimensional crossover in spin-orbit-coupled semiconductor nanowires
  with induced superconducting pairing.
\newblock {\em Phys. Rev. B}, 87:094518, Mar 2013.

\bibitem{Sun2012}
L.~Sun, L.~DiCarlo, M.~D. Reed, G.~Catelani, Lev~S. Bishop, D.~I. Schuster,
  B.~R. Johnson, Ge~A. Yang, L.~Frunzio, L.~Glazman, M.~H. Devoret, and R.~J.
  Schoelkopf.
\newblock Measurements of quasiparticle tunneling dynamics in a
  band-gap-engineered transmon qubit.
\newblock {\em Phys. Rev. Lett.}, 108:230509, Jun 2012.

\bibitem{Suominen2017}
H.~J. Suominen, M.~Kjaergaard, A.~R. Hamilton, J.~Shabani, C.~J. Palmstr\o{}m,
  C.~M. Marcus, and F.~Nichele.
\newblock Zero-energy modes from coalescing andreev states in a two-dimensional
  semiconductor-superconductor hybrid platform.
\newblock {\em Phys. Rev. Lett.}, 119:176805, Oct 2017.

\bibitem{Tessler2006}
R.~Tessler, C.~Saguy, O.~Klin, S.~Greenberg, E.~Weiss, R.~Akhvlediani,
  R.~Edrei, and A.~Hoffman.
\newblock Oxide-free \mbox{InSb} (100) surfaces by molecular hydrogen cleaning.
\newblock {\em Appl. Phys. Lett.}, 88:1918--031918, Jan 2006.

\bibitem{Thomas2019}
C.~Thomas, R.~E. Diaz, J.~H. Dycus, M.~E. Salmon, R.~E. Daniel, T.~Wang, G.~C.
  Gardner, and M.~J. Manfra.
\newblock Toward durable \mbox{Al-InSb} hybrid heterostructures via epitaxy of
  2\mbox{ML} interfacial \mbox{InAs} screening layers.
\newblock {\em Phys. Rev. Mater.}, 3:124202, Dec 2019.

\bibitem{vanVeen2018}
J.~van Veen, A.~Proutski, T.~Karzig, D.~I. Pikulin, R.~M. Lutchyn,
  J.~Nyg\aa{}rd, P.~Krogstrup, A.~Geresdi, L.~P. Kouwenhoven, and J.~D. Watson.
\newblock Magnetic-field-dependent quasiparticle dynamics of nanowire
  single-\mbox{C}ooper-pair transistors.
\newblock {\em Phys. Rev. B}, 98:174502, Nov 2018.

\bibitem{Vijay2016}
S.~Vijay and L.~Fu.
\newblock Teleportation-based quantum information processing with
  \mbox{M}ajorana zero modes.
\newblock {\em Phys. Rev. B}, 94:235446, 2016.

\bibitem{Whiticar2020}
A.~M. Whiticar, A.~Fornieri, E.~C.~T. O'Farrell, A.~C.~C. Drachmann, T.~Wang,
  C.~Thomas, S.~Gronin, R.~Kallaher, G.~C. Gardner, M.~J. Manfra, C.~M. Marcus,
  and F.~Nichele.
\newblock Coherent transport through a majorana island in an aharonov--bohm
  interferometer.
\newblock {\em Nat. Commun.}, 11(1):3212, Jun 2020.

\bibitem{winkler2019unified}
G.~W. Winkler, A.~E. Antipov, B.~Van~Heck, A.~A. Soluyanov, L.~I. Glazman,
  M.~Wimmer, and R.~M Lutchyn.
\newblock Unified numerical approach to topological
  semiconductor-superconductor heterostructures.
\newblock {\em Phys. Rev. B}, 99(24):245408, 2019.

\bibitem{Zimmermann1995}
U.~Zimmermann, S.~Abens, D.~Dikin, K.~Keck, and V.~M. Dmitriev.
\newblock Multiple andreev-reflection in asymmetric superconducting weak-links.
\newblock {\em Z. Phys. B}, 97(1):59--66, 1995.

\end{thebibliography}


\begin{thebibliography}{1}

\bibitem{Averin1995}
D.~Averin and A.~Bardas.
\newblock Ac \mbox{J}osephson effect in a single quantum channel.
\newblock {\em Phys. Rev. Lett.}, 75:1831--1834, Aug 1995.

\bibitem{Badawy2019}
G.~Badawy, S.~Gazibegovic, F.~Borsoi, S.~Heedt, C.-A. Wang, S.~Koelling, M.~A.
  Verheijen, L.~P. Kouwenhoven, and E.~P. A.~M. Bakkers.
\newblock High mobility stemless \mbox{InSb} nanowires.
\newblock {\em Nano Lett.}, 19(6):3575--3582, 2019.

\bibitem{Bardas1997}
A.~Bardas and D.~V. Averin.
\newblock Electron transport in mesoscopic disordered
  superconductor--normal-metal--superconductor junctions.
\newblock {\em Phys. Rev. B}, 56(14):R8518--R8521, October 1997.

\bibitem{Heedt2020}
S.~Heedt, M.~Quintero-P\'erez, F.~Borsoi, A.~Fursina, N.~van Loo, G.~P. Mazur,
  M.~P. Nowak, M.~Ammerlaan, K.~Li, S.~Korneychuk, J.~Shen, M.~A.~Y. van~de
  Poll, G.~Badawy, S.~Gazibegovic, K.~van Hoogdalem, E.~P. A.~M. Bakkers, and
  L.~P. Kouwenhoven.
\newblock Shadow-wall lithography of ballistic superconductor-semiconductor
  quantum devices.
\newblock {\em ArXiv e-prints}, 2007.14383, 2020.

\bibitem{nijholt2016orbital}
B.~Nijholt and A.~R. Akhmerov.
\newblock Orbital effect of magnetic field on the \mbox{M}ajorana phase
  diagram.
\newblock {\em Phys. Rev. B}, 93(23):235434, 2016.

\bibitem{Nowak2019}
M.~P. Nowak, M.~Wimmer, and A.~R. Akhmerov.
\newblock Supercurrent carried by nonequilibrium quasiparticles in a
  multiterminal \mbox{J}osephson junction.
\newblock {\em Phys. Rev. B}, 99(7):075416, February 2019.

\bibitem{Pan2020}
H.~Pan, W.~S. Cole, J.~D. Sau, and S.~Das~Sarma.
\newblock Generic quantized zero-bias conductance peaks in
  superconductor-semiconductor hybrid structures.
\newblock {\em Phys. Rev. B}, 101:024506, Jan 2020.

\bibitem{Tinkham1996}
M.~Tinkham.
\newblock {\em Introduction to superconductivity}.
\newblock Dover Publications, 1996.

\bibitem{vaitiekenas2020flux}
S.~Vaitiek{\.e}nas, G.~W. Winkler, B.~van Heck, T.~Karzig, M.-T. Deng,
  K.~Flensberg, L.~I. Glazman, C.~Nayak, P.~Krogstrup, R.~M. Lutchyn, and C.~M.
  Marcus.
\newblock Flux-induced topological superconductivity in full-shell nanowires.
\newblock {\em Science}, 367(6485), 2020.

\end{thebibliography}
\end{document}